\documentclass[cleveref,autoref,thm-restate]{jsat}
\usepackage{siunitx}
\usepackage{framed}
\usepackage{xspace}
\usepackage[table]{xcolor}

\pdfoutput=1 %uncomment to ensure pdflatex processing (mandatatory e.g. to submit to arXiv)
\hideJSAT  %uncomment to remove references to LIPIcs series (logo, DOI, ...), e.g. when preparing a pre-final version to be uploaded to arXiv or another public repository

\title{The QBF Gallery 2026}

\author{Cynthia Peyrer}{Institute for Symbolic Artificial Intelligence, Johannes Kepler University, Linz, Austria }{cynthia.peyrer@jku.at}{https://orcid.org/0009-0005-5751-9975}{}

\author{Luca Pulina}{Artificial Intelligence and Methods Lab, University of Sassari, Italy }{lpulina@uniss.it}{https://orcid.org/0000-0003-0258-3222}{}

\author{Martina Seidl}{Institute for Symbolic Artificial Intelligence, Johannes Kepler University, Linz, Austria }{martina.seidl@jku.at}{https://orcid.org/0000-0002-3267-4494}{}

\authorrunning{C. Peyrer and L. Pulina and M. Seidl}

\Copyright{Cynthia Peyrer and Luca Pulina and Martina Seidl}

\ccsdesc[500]{Theory of computation~Automated reasoning}

\keywords{Quantified Boolean Formula, QBF competition, QBF Gallery} 

\category{} %optional, e.g. invited paper

\relatedversion{} %optional, e.g. full version hosted on arXiv, HAL, or other respository/website
\supplement{\url{www.doi.org/10.5281/zenodo.22639316}}%optional, e.g. related research data, source code, ... hosted on a repository like zenodo, figshare, GitHub, ...
\funding{Parts of this work have been supported by the LIT AI Lab funded by the state of Upper
	Austria and by the Austrian Science Fund (FWF). [10.55776/COE12]}%optional, to capture a funding statement, which applies to all authors. Please enter author specific funding statements as fifth argument of the \author macro.

\acknowledgements{We thank Simone Heisinger for providing
her scripts and knowlege from the QBF Gallery 2023 to this 
project.}%optional

\nolinenumbers %It is recommended to enable line numbers for submissions via 'HAL' or 'Zenodo'.

\EventNoEds{2}
\EventLongTitle{This article is submitted to JSAT}
\EventShortTitle{JSAT YYYY}
\EventAcronym{JSAT}
\EventYear{YYYY}
\EventDate{}
\EventLocation{Little Whinging, United Kingdom}
\EventLogo{}
\SeriesVolume{VV(I))} % Need to match!
\ArticleNo{XX(Y)} % Need to match!
\Submitted{MMMM YYYY}
\Revised{MMMM YYYY}
\newcommand{\qbf}{\ensuremath{\phi}}
\newcommand{\prefix}{\ensuremath{\Pi}}
\newcommand{\propmatrix}{\ensuremath{\Psi}}

\begin{document}

\maketitle

\begin{abstract}
The QBF Gallery 2026 is the latest evaluation event of (dependency) quantified Boolean formula 
(QBF) solving technology.
Its goal is to assess the current state-of-the-art, enable the evaluation of novel solvers on
newly curated benchmark instances and present the progress made since the previous competition.
This report discusses the solvers, instances and tools submitted by the community and
introduces new benchmark collections for evaluation. 
We evaluate solvers in three tracks, reflecting most
common input formats in (dependency) QBF solving.
In addition to those main evaluations, we investigate effects of preprocessing
and include a track of new crafted benchmark families.

The results are analyzed using various metrics and complementary statistics
on solver similarities, different memory limitations and performance across formula 
families.
Finally, the dedicated benchmarks for each track are made available as an artifact to support
future research and reproducible evaluation.
\end{abstract}

\section{Introduction}
Quantified Boolean formulas (QBFs) arise in numerous practical applications
including synthesis, bounded model checking and planning ~\cite{shukla2019survey}. 
The corresponding decision problem is widely regarded as the canonical PSPACE-complete
problem~\cite{stockmeyer1973word}. 
To assess the state-of-the-art in QBF solving, regular events such as
the QBFEval and the QBF Gallery have been organized.
This tradition began in 2003 and had its latest entry 2023, with numerous editions in 
between~\cite{berre2004second, berre2003challenges, heisinger2026qbfgallery2023, janota2014qbfgallery, lonsing2015qbf, narizzano2006report, peschiera2010seventh, pulina2016ninth, pulina2019qbf}.
This paper reports on the organization and results of the QBF Gallery 2026.

The event itself was conducted virtually with the evaluations performed on a local compute cluster.
The results were presented at the Federated Logic Conference (FLoC) 2026,
as part of the FLoC Olympics, affiliated with the \emph{International Conference on Theory and 
Applications of Satisfiability Testing}~\cite{sat2026}.
In addition to solvers and preprocessors, participants were invited to submit
novel problem instances to different tracks covering different QBF formalisms and input formats.
The results are analyzed not only to determine a winner for each track 
but also to identify the practical challenges currently faced by the field.
By including instances and solvers from previous competitions, we aim to show the
progress made over the last iterations.
This paper is structured as follows:
First, we describe the organizational aspects of the Gallery including the 
experimental setup and the individual tracks.
We then present the benchmarks and solvers newly submitted and further the 
selected participants of this Gallery.
Finally, we evaluate the results on their different performance metrics and provide a deeper
analysis of the collected data.

\section{Organization}

The QBF Gallery 2026 featured five tracks, each based on different types of formulas.
This section describes the experimental setup including resource limits and evaluation
procedure as well as the input formats and semantic details of of each track.

\subsection{Setup}
All experiments were conducted on an Ubuntu 24.04.4 LTS compute cluster 
consisting of 20 nodes, each equipped
with 64 dual-socket 16-core AMD EPYC 7313@3.7 GHz CPUs.
All tracks were evaluated on this setup with one
allocated node per instance.
A timeout of \SI{900}{\second} and a memory limit
of \SI{8}{\giga\byte} was set for each instance 
in every track.
Additionally, some selected tracks were re-ran with an increased
memory limit of \SI{128}{\giga\byte} to evaluate the extent to which solver
performance is constrained by memory limitation. 

\subsection{Tracks of the QBF Gallery 2026}
In the following we present the Prenex Conjunctive Normal Form (PCNF),
Prenex Non-Conjunctive Normal Form (PNCNF), Dependency Quantified Boolean Formula (DQBF), Crafted Instances and
Preprocessor track.
Submissions of both formulas and solvers were accepted for the first four tracks.
In contrast, the Preprocessor track only accepted preprocessors while using the 
benchmarks submitted to the PCNF (resp. Non-PCNF) track.
Solvers and formulas were allowed to be featured in multiple tracks, provided they 
satisfy the respective track requirements.

\newcommand{\qdimacs}{\texttt{QDIMACS}\xspace}

\begin{figure}[!b]
	\begin{minipage}{.3\linewidth}
		\begin{framed}
			\noindent\texttt{p cnf 2 2}\\
			\texttt{a 1 0}\\
			\texttt{e 2 0}\\
			\texttt{1 2 0}\\
			\texttt{-1 -2 0}
		\end{framed}	
	\end{minipage}
	\caption{Example of a \qdimacs representation of $\forall a \exists x: (a \vee x) \wedge (\neg a \vee \neg x)$}
	\label{fig:qdimacs}
\end{figure}

\subsubsection*{PCNF Track}
A QBF in Prenex Conjunctive Normalform (PCNF) has the structure
$\qbf = \prefix : \propmatrix$, where
$\prefix = Q_1 x_1 \dots Q_m x_m$ is a quantifier prefix with 
$Q_i \in \{\forall, \exists\}$
and $\propmatrix$ is a propositional matrix in conjunctive normal form, i.e.,
a conjunction of clauses (disjunctions) of literals over pairwise distinct
propositional variables $x_1 \dots x_m$.
The standardized input format for PCNFs---and consequently the required format
of this track---is \qdimacs~\cite{qdimacs}.
Figure~\ref{fig:qdimacs} shows an example \qdimacs encoding of an exclusive-or formula
eligible for the PCNF track.

\newcommand{\qcir}{\texttt{QCIR}\xspace}
\subsubsection*{PNCNF Track}
In contrast to the PCNF track, the prenex non-CNF (PNCNF) track permits arbitrary 
propositional formulas as the matrix $\propmatrix$.
The requirements regarding the quantifier prefix are the same as in the PCNF track.
Input formulas for this track are accepted in the \qcir (Quantified CIRcuit)~\cite{qcir}
format.
The left side of figure~\ref{fig:qcir} shows an example of an \qcir-encoded QBF.
A \qcir formula consists of a prefix as well as a sequence of gate definitions bound 
to Tseitin-like variables~\cite{tseitin1983complexity} called \emph{gate variables}
forming a directed acyclic graph (DAG).
Supported gates include \texttt{and}, \texttt{or}, \texttt{xor} (exclusive-or), 
\texttt{ite} (if-then-else) as well as local quantifiers \texttt{forall} and 
\texttt{exists}.
Consequently, \qcir formulas can represent non-prenex QBFs.
To ensure that submitted formulas are in PNCNF, the Gallery imposes the following
restrictions:
\begin{itemize}
	\item No free variables (closed prefix)
	\item Only \texttt{and} and \texttt{or} gates are permitted
	\item Local quantifier gates are not permitted
	\item Variables may use numeric names only
\end{itemize}
The resulting restricted format is commonly referred to as 
\texttt{cleansed QCIR}, of which an example is shown on the right side of 
figure~\ref{fig:qcir}. 
The semantics of submitted formulas in this format 
were checked using dedicated tools during benchmark selection.

\begin{figure}[!t]
	\begin{minipage}{.4\linewidth}
		\begin{framed}
			\noindent\texttt{\#QCIR-14}\\
			\texttt{free(z)}\\
			\texttt{forall(a,b)}\\
			\texttt{exists(x)}\\
			\texttt{output(g4)}\\
			\texttt{g1 = xor(a,-y)}\\
			\texttt{g2 = exists(y; g1)}\\
			\texttt{g3 = and(z, x)}\\
			\texttt{g4 = ite(b, g3, g2)}
		\end{framed}	
	\end{minipage}%
	\begin{minipage}{.4\linewidth}
		\begin{framed}
			\noindent\texttt{\#QCIR-14}\\
			\texttt{forall(1,2,3)}\\
			\texttt{exists(4,5,6)}\\
			\texttt{output(11)}\\
			\texttt{7 = or(-1, 4)}\\
			\texttt{8 = or(-2, 5)}\\
			\texttt{9 = or(-3, 6)}\\
			\texttt{10 = and(4, 5, 6)}\\
			\texttt{11 = and(7, 8, 9, -10)}
		\end{framed}	
	\end{minipage}
	\caption{Example of a \qcir (left) representing $\forall a \forall b \exists x: \texttt{ite}(b, z \wedge x, \exists y: \texttt{xor}(a, \neg y))$ and cleansed \qcir (right) instance representing
	$\forall a \forall b \forall c \exists x \exists y \exists z: (\neg a \vee x) \wedge (\neg b \vee y) \wedge 
	(\neg c \vee z) \wedge \neg (x \wedge y \wedge z)$
	} 
	\label{fig:qcir}
\end{figure}

\newcommand{\dqdimacs}{\texttt{DQDIMACS}\xspace}

\subsubsection*{DQBF Track}
Dependency QBF (DQBF) extends QBF by allowing a non-linear quantifier prefix \prefix. 
Instead of a variable being dependent on their predecessor in a linear list, 
variable dependencies are stated explicitly utilizing Henkin
quantifiers~\cite{balabanov2014henkin}.
This increased expressiveness also raises the complexity of the decision problem to solve them 
from PSPACE to NEXPTIME.
The input format for this track is \dqdimacs~\cite{dqdimacs}, an extension of 
\qdimacs additionally supporting the definition of Henkin quantifiers and their
associated dependencies.
Figure~\ref{fig:dqdimacs} shows a \dqdimacs encoding of a false exclusive-or formula, 
which can be constructed by using explicit dependencies.

\begin{figure}[!t]
	\begin{minipage}{.3\linewidth}
		\begin{framed}
			\noindent\texttt{p cnf 4 2}\\
			\texttt{a 1 2 0}\\
			\texttt{d 3 1 0}\\
			\texttt{d 4 2 0}\\
			\texttt{1 4 0}\\
			\texttt{-1 -4 0}
		\end{framed}	
	\end{minipage}
	\caption{Example of a \dqdimacs representation of $\forall a \forall b \exists x(a) \exists y(b): (a \vee y) \wedge (\neg a \vee \neg y)$}
	\label{fig:dqdimacs}
\end{figure}

\subsubsection*{Crafted Instances Track}
Crafted instances are parameterized formula families originating from proof complexity,
such as the Kleine B\"{u}ning--Karpinski--Fl\"{o}gel (KBKF) 
formulas~\cite{buning1995resolution}.
They are designed to assess the strength of the underlying proof systems of QBF solvers.
Since the size of these instances can be adjusted, they enable the analysis of how solving effort increases 
with increasing instance parameter.
This scalability analysis is the primary objective of the Crafted Instances track.
As all submitted crafted instances are in PCNF format, this track features the 
corresponding solvers.

\subsubsection*{Preprocessor Tracks}
The Preprocessor track is the only track that does not accept formulas or solvers for submission.
Instead, it focuses on evaluating preprocessors. 
A preprocessor rewrites the input into an equivalent formula without the primary goal of solving it but rather 
making it (potentially) easier for a subsequent solver to handle.
Although also widely employed in inprocessing, preprocessors are executed independently on the benchmarks
of a track resulting in a set of novel instances to evaluate.
This track features the formulas of the PCNF and PNCNF tracks.
No preprocessor was submitted for the DQBF track.

\section{Benchmarks}
In the following section, we present the instances submitted to and 
selected for each track.
In addition to describing their origins, we provide basic information about 
the problems they encode and their structural characteristics.
The selection process for the instances was the same for all tracks. 
First, all newly submitted formulas were solved using the eligible solvers 
for the respective track.
Each run in this \emph{smoke test} was subject to a time limit of \SI{300}{\second} with four 
instances executed concurrently on a single node.
Instances solved by every solver were removed from the 
selection pool. 
From the remaining formulas, we randomly chose a representative subset.
To complete the benchmarks, we selected a subset of instances from
the previous QBF Gallery 2023~\cite{heisinger2026qbfgallery2023}.
For the DQBF track, we chose instances from the QBFEval 
2020~\cite{qbfeval20}, as the same set of formulas was used in 2023.
Finally, we compared the results reported by the solvers to identify 
disagreements and thus potential unsoundness.
The selected formulas of the tracks can be found in the
artifact published on Zenodo~\cite{gallery26-instances}.

\begin{table}[!t]
	\scriptsize
	\centering
	\begin{tabular}{|l|l|c|c|c|c|}
		\hline	
		\textbf{Family} & \textbf{Author} & \textbf{Sub. (Sel.)} & \textbf{sat} & \textbf{unsat} & \textbf{unsolved} \\
		\hline
		Bounded Synthesis & Slivovsky F. & 115 (40) & 41 (16) & 39 (12) & 35 (12)\\
		\hline
		Connect4 & Mayer-Eichberger V. et al & 170 (64) & 42 (13) & 101 (45) & 27 (6)\\
		\hline
		Feature Models & Heisinger S. & 20 (16) & 2 (2) & 9 (5) & 9 (9) \\
		\hline
		GDL-Games & He Y. et al. & 152 (68) & 42 (21) & 73 (38) & 37 (9)\\
		\hline
		Pasp2QBF & Mau\'a et al. & 66 (30) & 44 (23) & 22 (7) & 0 (0) \\
		\hline
		\hline
		\textbf{Total} & & 523 (218) & 171 (75) & 244 (107) & 108 (36)\\
		\hline
	\end{tabular}
	\caption{Submitted (Selected) QDIMACS formulas of the PCNF track}
	\label{tbl:pcnf-overview}
\end{table}

\subsection{PCNF Benchmarks (QDIMACS Format)}
Table~\ref{tbl:pcnf-overview} provides an overview of the submitted and 
selected formulas for the PCNF track, while detailed statistical data
of the selected formulas is summarized in Table~\ref{tbl:pcnf-stats}.
In total, we received 523 instances across five different submissions 
of which 218 were selected for the Gallery. 
Additionally, we took 88 instances from the QBF Gallery
2023~\cite{heisinger2026qbfgallery2023}, resulting in a total of 306 
instances for the PCNF dataset. 
In the following we describe the submitted formula families.

\begin{table}[!t]
	\scriptsize
	\centering
	\begin{tabular}{|l|c|c|c|c|c|c|}
		\hline
		\textbf{Dataset} & \textbf{Q. Alt.} & \textbf{Avg. Clauses} & \textbf{Avg. Vars} & \textbf{Avg. $\forall$.} & \textbf{Avg. $\exists$} & \textbf{\#} \\
		\hline
		Bounded Synthesis & 2 & 219195 & 62034 & 5 & 62029 & 40\\
		\hline
		Connect4: partial7x6 & 6-18 & 9043 & 2041 & 20 & 2020 & 30\\
		\hline
		Connect4: fromEmpty & 10-24 & 9034 & 2010 & 26 & 1983 & 34\\
		\hline
		Feature Models & 1 & 617841 & 428237 & 9995 & 418242 & 16\\
		\hline
		GDL-Games: GB & 32-72 & 46326 & 13115 & 95 & 13020 & 17\\
		\hline
		GDL-Games:  GB (bloqqer) & 12-48 & 9205 & 15496 & 38 & 1026 & 15\\
		\hline		
		GDL-Games: GD & 24-72 & 56361 & 19068 & 99 & 18969 & 19\\
		\hline		
		GDL-Games: GD (bloqqer) & 10-34 & 14416 & 19301 & 37 & 1862 & 17\\
		\hline
		Pasp2QBF & 1 & 6037 & 1644 & 139 & 1505 & 30\\				
		\hline
		\hline
		Gallery23: Matrix Multiplication & 2 & 305902 & 44962 & 24 & 44938 & 1\\
		\hline
		Gallery23: Grid Games & 12-20 & 4840 & 1671 & 53 & 1618 & 6\\		
		\hline
		Gallery23: QBFEval 2020 & 1-44 & 267796 & 83839 & 2123 & 62572 & 62\\
		\hline
		Gallery23: Hex Games & 12 & 7577 & 1999 & 32 & 1968 & 4\\
		\hline
		Gallery23: Organic Synthesis & 2 & 136281 & 38860 & 14 & 38846 & 4\\
		\hline
		Gallery23: Pattern Finding & 2 & 14695 & 6232 & 83 & 6149 & 8\\
		\hline
		Gallery23: Linear Domino & 11-12 & 6359 & 1115 & 140 & 975 & 3\\
		\hline				
	\end{tabular}
	\caption{Statistics of the selected formulas in the PCNF track}
	\label{tbl:pcnf-stats}
\end{table}

\begin{itemize}
	\item The \emph{Bounded Synthesis} instances encode the question of whether a Mealy machine of at most $k$ states realizes or falsifies a given LTL specification. 
	The original encoding was introduced by Faymonville et al.~\cite{faymonville2017encodings}.
		
	\item The \emph{Connect4} instances~\cite{connect4-instances} encode
	the popular \emph{Connect Four} game as QBFs.
	Depending on the initial state of the game, the formulas are divided
	into two subfamilies.
	\emph{Partial7x6} instances begin from an ongoing 
	game whereas \emph{fromEmpty} instances assume
	an empty board.
	The \texttt{QDIMACS} instances were exported using Bule~\cite{jung2022qbf}.
		
	\item The \emph{Feature Models} instances~\cite{featuremodels-instances} 
	encode feature-model differences as introduced by Heisinger 
	et al.~\cite{heisinger2025semantic}.
	They include configurations ranging from bicycles to the Linux kernel.
	Initially, the problems were encoded in Unified Variability Language 
	(UVL)~\cite{sundermann2021yet} and afterwards
	exported by Booleguru~\cite{heisinger2024booleguru}.	
	
	\item The \emph{GDL-Games}~\cite{gdlgames-instances} instances are 
	derived from a general Game Description Language (GDL)
	encoding.
	They express whether a player can achieve 100 points within $n$ steps, 
	regardless of the actions of the other players.
	Details of the encoding were published by He et al.~\cite{he2024solving}.
	The benchmark set is divided into four sub-families:
	\emph{GB} instances use a baseline encoding while \emph{GD} instances
	use a dependency-based encoding.	
	Both sets were additionally preprocessed using 
	bloqqer~\cite{biere2011bloqqer}, resulting in the
	\emph{GD (bloq)} and \emph{GB (bloq)} subfamilies.
	
	\item The \emph{Pasp2QBF} instances~\cite{pasp2qbf-instances}
	encode probabilistic answer set programming (PASP) inference 
	problems as 2QBF formulas~\cite{wang2024compositional}. 
	The collection contains 12 sub-families representing different 
	problems ranging from Sudoku puzzles to graph coloring. 
	To generate the instances, the PASP problems were first encoded in WCNF 
	formulas and subsequently rewritten as QBFs.
			
	\item To complete the PCNF track, we also included instances  
	from the \emph{QBF Gallery 2023}~\cite{gallery23-instances}.
	Detailed descriptions of the formula families can be found in the 
	report of the Gallery 2023~\cite{heisinger2026qbfgallery2023}.
\end{itemize}

\subsection{PNCNF Benchmarks (QCIR Format)}
Table~\ref{tbl:nonpcnf-overview} provides an overview of the submitted and 
selected formulas for the PNCNF track, while detailed statistical data of 
the selected formulas is shown in Table~\ref{tbl:nonpcnf-stats}.
In total, we received 554 instances across two submissions of which 41 
were selected for the Gallery.
Additionally, we took 335 instances from the QBF Gallery
2023~\cite{heisinger2026qbfgallery2023} resulting in a total of 376
instances for the PNCNF dataset.
In the following, we describe the submitted formula families.

\begin{table}[!t]
	\scriptsize
	\centering
	\begin{tabular}{|l|l|c|c|c|c|}
		\hline	
		\textbf{Family} & \textbf{Author} & \textbf{Sub. (Sel.)} & \textbf{sat} & \textbf{unsat} & \textbf{unsolved} \\
		\hline
		Feature Models & Heisinger S. & 20 (16) & 0 (0) & 6 (2) & 14 (14)  \\
		\hline
		Functional Synthesis & Khalimov A. & 334 (25) & 311 (20) & 0 (0) & 23 (5)  \\
		\hline
		\hline
		\textbf{Total} & & 554 (41) & 311 (20) & 6 (2) & 37 (19)\\
		\hline		
	\end{tabular}
	\caption{Submitted (Selected) formulas of the PNCNF track}
	\label{tbl:nonpcnf-overview}
\end{table}

\begin{itemize}	
	\item The \emph{Feature Models} instances are the same as described
	for the PCNF track, but exported from the original UVL 
	into the \texttt{QCIR} format using 
	the tool Booleguru~\cite{heisinger2024booleguru}.
		
	\item The \emph{Functional Synthesis} 
	instances~\cite{functionalsynthesis-instances}
	encode the automatic synthesis of a system from a behavioral 
	specification.
	The resulting instances are formulated as safety games.
	
	\item To complete the PNCNF track, we also included instances from 
	the \emph{QBF Gallery 2023}~\cite{gallery23-instances}.
	Detailed descriptions of the formula families can be found in the report of 
	the QBF Gallery 2023~\cite{heisinger2026qbfgallery2023}.
\end{itemize}

\begin{table}[!t]
	\scriptsize
	\centering
\begin{tabular} {|l|c|c|c|c|c|c|}
	\hline
	\textbf{Dataset} & \textbf{Q. Alt.} & \textbf{Avg. gates} & \textbf{Avg. vars} & \textbf{Avg. $\forall$} & \textbf{Avg. $\exists$} & \textbf{\#}\\
	\hline
	Feature Models & 1 & 50144 & 19990 & 9995 & 9995 & 16\\
	\hline
	Functional Synthesis & 1 & 243962 & 193 & 186 & 7 & 25\\
	\hline
	\hline
	Gallery23: Grid Games & 10-20 & 1293 & 140 & 47 & 94 & 19\\
	\hline
	Gallery23: Transitive Closure & 36-110 & 20310 & 4290 & 2884 & 1406 & 10\\
	\hline
	Gallery23: Nested Counterfactuals & 7 & 904 & 187 & 96 & 91 & 6\\
	\hline
	Gallery23: Generalized TTT & 6-14 & 847 & 67 & 21 & 46 & 28\\
	\hline
	Gallery23: Hyper-Properties & 0-1 & 5071952 & 1460 & 592 & 868 & 20\\
	\hline
	Gallery23: Matrix Multiplication & 2 & 15103 & 1864 & 25 & 1839 & 10\\
	\hline
	Gallery23: Hadwiger Conjecture & 1 & 28849 & 196 & 118 & 78 & 21\\
	\hline
	Gallery23: QBFEval 2020 & 1-7 & 41344 & 2141 & 1185 & 956 & 191\\
	\hline
	Gallery23: Circuit Minimization & 1-2 & 14036 & 403 & 193 & 211 & 30\\
	\hline
\end{tabular}
\caption{Statistics of the selected formulas in the PNCNF track}
\label{tbl:nonpcnf-stats}
\end{table}

\subsection{DQBF Benchmarks (DQDIMACS Format)}
Table~\ref{tbl:dqbf-overview} provides an overview of the submitted
and selected formulas for the DQBF track, while detailed statistical 
data of the selected formulas is shown in Table~\ref{tbl:dqbf-stats}.
In total, we received 980 instances across two submissions of which 139 
were selected for the gallery.
Additionally, we took 181 instances from the QBFEval 2020~\cite{qbfeval20}
resulting in a total of 320 instances for the DQBF dataset.
In the following we describe the submitted formula families.

\begin{table}[!t]
	\scriptsize
	\centering
	\begin{tabular}{|l|l|c|c|c|c|}
		\hline	
		\textbf{Family} & \textbf{Author} & \textbf{Sub. (Sel.)} & \textbf{sat} & \textbf{unsat} & \textbf{unsolved} \\
		\hline
		Bounded Synthesis & Slivovsky F. & 386 (73) & 128 (16) & 87 (26) & 171 (31)\\	
		\hline
		Ramsey & Cheng C. & 504 (44) & 241 (18) & 31 (6) & 232 (20)\\	
		\hline
		Succinct Graph & Cheng C. & 90 (22) & 10 (5) & 42 (9) & 38 (8)\\	
		\hline
		\hline
		\textbf{Total} & & 980 (139) & 379 (39) & 160 (41) & 441 (59) \\
		\hline
	\end{tabular}
	
	\caption{Submitted (Selected) formulas of the DQBF track}
	\label{tbl:dqbf-overview}
\end{table}

\begin{itemize}	
	\item The \emph{Bounded Synthesis} instances follow the same encoding 
	as its PCNF-track counterpart but additionally uses 
	Henkin dependencies. 
	Two subfamilies were submitted:
	the \emph{state-symbolic} encoding has propositional
	inputs whereas the \emph{fully-symbolic} encoding has universal 
	inputs, making it the most concise of the variants.
			
	\item The \emph{Ramsey} and \emph{Succinct Graph} 
	instances~\cite{ramseysucc-instances} were submitted
	together with their generator by the same author as one single
	submission of two formula families.
	The \emph{Ramsey} formulas check if a Ramsey 
	number~\cite{ajtai1980note} is greater
	than some threshold.
	The \emph{Succinct Graph} formulas encode graph problems 
	from the three subcategories: 
	\emph{Clique}, \emph{Coloring} and \emph{Hamiltonian}.
	
	\item To complete the DQBF track, we also included instances from the 
	\emph{QBFEval 2020}~\cite{qbfeval20}.
\end{itemize}

\begin{table}[!t]
\scriptsize
\centering
\begin{tabular} {|l|c|c|c|c|c|c|c|}
	\hline
	\textbf{Dataset} & \textbf{Q. Alt.} & \textbf{Avg. clauses} & \textbf{Avg. vars} & \textbf{Avg. $\forall$} & \textbf{Avg. $\exists$} & \textbf{Avg. $d$} & \textbf{\#}\\
	\hline
	Bounded Synthesis: Full & 1 & 2649624 & 100672 & 23 & 100636 & 14 & 36\\		
	\hline
	Bounded Synthesis: State & 1 & 2086759 & 581633 & 12 & 575400 & 6220 & 37\\
	\hline
	Ramsey & 1 & 389 & 126 & 24 & 90 & 12 & 44\\
	\hline
	Succinct Graph: Clique & 1 & 951 & 247 & 6 & 173 & 68 & 8\\
	\hline
	Succinct Graph: Hamiltonian & 1 & 293 & 2108 & 17 & 2074 & 17 & 10\\		
	\hline
	Succinct Graph: Coloring & 1 & 172 & 61 & 24 & 32 & 5 & 4\\		
	\hline
	\hline
	QBF Eval 2020 & 0-2 & 4105 & 921 & 89 & 747 & 85 & 181\\
	\hline
\end{tabular}
\caption{Statistics of the selected formulas in the DQBF track}
\label{tbl:dqbf-stats}	
\end{table}

\subsection{Crafted Instances}
Table~\ref{tbl:crafted-overview} provides an overview of
the submitted and selected formulas for the Crafted Instances track. 
We received one submission (\texttt{QBFFamExt}~\cite{qbffamext}) containing six 
new families all of which were selected for the track.
Each family is parameterized by $n$ and some families provide 
additional parameters for further tuning.
Using these parameters, we selected 16 instances across the six families.
The details of this selection are shown in
Table~\ref{tbl:crafted-selected}.
In the following, we describe the selected formula families.

\begin{table}[!b]
	\scriptsize
	\centering
	\begin{tabular}{|l|l|l|c|}
		\hline	
		\textbf{Family} & \textbf{Abbr.} & \textbf{Author} & \textbf{result} \\
		\hline
		Gadget-Family & \texttt{GADGET} & A. Schleitzer, O. Beyersdorff & unsat \\
		\hline
		All-Equal-SAT & \texttt{AESAT} & A. Schleitzer, O. Beyersdorff & unsat \\
		\hline		
		Succinct-K-Radius & \texttt{SUCCKRAD} & A. Schleitzer, O. Beyersdorff &  unsat \\	
		\hline
		Clique Coloring & \texttt{CLIQUECOL} & A. Schleitzer, O. Beyersdorff & unsat\\	
		\hline		
		Generalised Subset Sum & \texttt{SUBSETSUM} & A. Schleitzer, O. Beyersdorff & unsat \\	
		\hline		
	\end{tabular}
	
	\caption{New Crafted Instance formulas}
	\label{tbl:crafted-overview}
\end{table}

\begin{itemize}	
	\item The \emph{All-Equal-SAT} family~\cite{schleitzer2025computationally} encodes the following problem:
	Given a 3-CNF formula $\varphi(X,Y)$, where $(X,Y)$ is a partition 
	of $\mathrm{vars}({\varphi})$,
	All-Equal-$\exists\forall$-3SAT solves, whether there exists an 
	assignment to the variables in $X$ such that 
	for every assignment to the variables in $Y$, at least one 
	clause contains either only true or only false literals.
	The parameter $n$ specifies the number of clauses in the generated
	formula.
	
	\item The \emph{Clique Coloring} 
	family~\cite{schleitzer2025computationally} encode the decision problem
	given a graph $G=(V,E)$ and $k=3$, whether $G$ admits a 
	$k$-clique-coloring, 
	i.e. a coloring of the vertices in $V$ such that no maximal clique in 
	$G$ is monochromatic.
	The parameter $n$ specifies the family index of the Gr\"{o}tzsch graph induction.
	
	\item The \emph{Gadget-Family}~\cite{schleitzer2023classes} problems 
	are obtained by connecting a propositional base formula with a gadget.
	Available base formulas are \emph{single clause} (SC), 
	\emph{implication chain} (IC) and \emph{equality chain} (EC).
	The gadget can be chosen to be either equality (EQ) or exclusive-or 
	(XOR) constraints.
	Both the base formula and the gadget are additional parameters of the 
	construction.
	The parameter $n$ specifies either the number of variables for SC or 
	the chain length for IC and EC.	
	
	\item The \emph{Generalized Subset Sum}~\cite{schleitzer2025computationally} family
	encodes the following problem:
	Given two integer vectors $u$ and $v$ with $|u|=k$, $|v|=l$ and an 
	integer $t$, the formula encodes 
	$\exists x\forall y \colon (ux+vy\neq t)$, 
	where $x$ and $y$ are binary vectors with $|x|=k$, $|y|=l$.	
	The parameter $n$ scales the lengths of $u$ and $v$ and determines
	the number of selector variables in the encoding.
	
	\item The \emph{Succinct-K-Radius} family encodes whether a 
	Galperin-Widgerson representation $C$ of a directed 
	graph $G=(V_G,E_G)$ has a radius of at most $k$.
	The value $k$ is an additional parameter of the construction. 
	The parameter $n$ specifies the grid height.
\end{itemize}

\begin{table}[!t]
	\scriptsize
	\centering
	\begin{tabular}{|l|l|c|}
		\hline	
		\textbf{Instance} & \textbf{range} & \textbf{\#} \\
		\hline
		\texttt{AESAT(n)} & $n=5\dots30$, 40,50,60,70 & 30 \\
		\hline		
		\texttt{CLIQUECOL(n)} & $n=1\dots4$ & 4 \\	
		\hline		
		\texttt{GADGET(n, base=ec, gadget=eq)} & $n=5\dots30$, 40,50,60,70 & 30 \\
		\hline
		\texttt{GADGET(n, base=ec, gadget=xor)} & $n=5\dots30$, 40,50,60,70 & 30 \\
		\hline
		\texttt{GADGET(n, base=ic, gadget=eq)} & $n=5\dots30$, 40,50,60,70 & 30 \\
		\hline
		\texttt{GADGET(n, base=ic, gadget=xor)} & $n=5\dots30$, 40,50,60,70 & 30 \\
		\hline
		\texttt{GADGET(n, base=sc, gadget=eq)} & $n=5\dots30$, 40,50,60,70 & 30 \\
		\hline
		\texttt{GADGET(n, base=sc, gadget=xor)} & $n=5\dots30$, 40,50,60,70 & 30 \\
		\hline
		\texttt{SUBSETSUM(n)} & $n=5\dots30$, 40,50,60,70 & 30 \\	
		\hline				
		\texttt{SUCCKRAD(n,k=3)}  & $n=5\dots30$, 40,50,60,70 & 30 \\
		\hline		
		\texttt{SUCCKRAD(n,k=5)}  & $n=5\dots30$, 40,50,60,70 & 30 \\
		\hline		
		\texttt{SUCCKRAD(n,k=7)}  & $n=5\dots30$, 40,50,60,70 & 30 \\
		\hline		
		\texttt{SUCCKRAD(n,k=10)}  & $n=5\dots30$, 40,50,60,70 & 30 \\
		\hline		
		\texttt{SUCCKRAD(n,k=15)}  & $n=5\dots30$, 40,50,60,70 & 30 \\
		\hline		
		\texttt{SUCCKRAD(n,k=20)}  & $n=5\dots30$, 40,50,60,70 & 30 \\
		\hline		
		\texttt{SUCCKRAD(n,k=30)}  & $n=5\dots30$, 40,50,60,70 & 30 \\
		\hline		
	\end{tabular}
	
	\caption{Selected formulas of the Crafted Instances track}
	\label{tbl:crafted-selected}
\end{table}

\section{Solvers and Preprocessors}
In the following, we introduce the solvers and preprocessors participating 
in each track.
The participants include not only newly submitted tools but also solvers
and preprocessors introduced in previous QBF Galleries.
Some solvers appear in multiple tracks because of their support for different 
input formats.

\newcommand{\caqe}{\texttt{caqe}\xspace}
\newcommand{\depqbf}{\texttt{DepQBF}\xspace}
\newcommand{\dynqbf}{\texttt{DynQBF}\xspace}
\newcommand{\hybridqbf}{\texttt{HybridQBF}\xspace}
\newcommand{\miniqu}{\texttt{MiniQU}\xspace}
\newcommand{\qute}{\texttt{Qute}\xspace}
\newcommand{\rareqs}{\texttt{RAReQS}\xspace}

\subsection{PCNF Solvers}
Table~\ref{tbl:pcnf-solver-overview} lists the solvers participating in the PCNF track.
Three newly submitted solvers, each with up to 8 different configurations, were included. 
The remaining solvers were selected from the previous QBF Gallery.

\begin{itemize}
	\item \hybridqbf is based on \depqbf and incorporates additional solving techniques including 
	WalkSAT~\cite{wei2004towards} and glue clauses~\cite{audemard2009predicting}.
	No accompanying paper or system description is available.
	The submitted versions utilize different configurations involving the solver \caqe~\cite{rabe2015caqe} as well 
	as \texttt{Bloqqer}~\cite{biere2011bloqqer}
	and \texttt{HQSPre}~\cite{wimmer2017hqspre} for pre- and inprocessing.
	
	\item \texttt{\depqbf-cadical}~\cite{depqbf-cadical} replaces the internal SAT solvers
	of the original \depqbf implementation, \texttt{Minisat}~\cite{minisat}, for \texttt{CaDiCaL}~\cite{biere2024cadical}.
	
	\item \qute~\cite{peitl2019dependency} was submitted in three different configurations:
	one using out-of-order (ofo) decisions~\cite{bohm2024should}, one implementing the $D_{\forall \text{pure}}$ (upure) 
	dependency scheme~\cite{chew2026strong} and the (plain) standard configuration~\cite{peitl2019dependency}.	
\end{itemize}

\noindent In the following we list the different solving paradigms applied by the PCNF
solvers:

\begin{itemize}
	\item \textbf{CEGAR} (Counter Example Guided Abstraction Refinement) iteratively 
	solves propositional abstractions of a QBF formula~\cite{Clarke2003Counterexample}.
	Candidate results obtained from an abstraction are validated against the original 
	formula and refined. 
	This process is repeated until the QBF is determined to be true or false.	
	CEGAR is commonly seen as a two player game between the universal
	and existential player, trying to find a winning strategy for either player.
	The participating PCNF solvers utilize two flavours of CEGAR:
	\begin{itemize}
		\item \textbf{CEGAR} with \textbf{clausal abstraction} decomposes the formula
		into an abstraction of a sequence of propositional formulas 
		associated with the quantifier blocks~\cite{rabe2015caqe}.
				
		\item \textbf{CEGAR} with \textbf{expansion} recursively\footnote{A 
			non-recursive version of this approach has
			been introduced~\cite{bloem2018expansion}, 
			but is not applied in any participating solver} 
		expands quantified variables into propositional instances.
		(Counter)-models discovered during solving are then used to 
		refine the expansion until a result is obtained~\cite{janota2016solving}.		
	\end{itemize}

	\item \textbf{Dynamic programming} first partitions the CNF matrix into 
	sub-problems using a tree decomposition approach and then processes this decomposition by dynamic 
	programming~\cite{charwat2016dynamic}, combining intermediate results from 
	the subproblems.
	To account for dependencies induced by the quantifier prefix, dependency schemes are 
	integrated into the solving process.
	This approach can be viewed as an expansion-based method as it combines 
	local expansions of the subproblems.	
	
	\item \textbf{QCDCL} (Quantified Conflict Driven Clause learning) extends CDCL-based SAT solving by
	learning conflict clauses and solution cubes.
	Once an empty constraint can be learnt, the result is determined~\cite{beyersdorff2021quantified}. 	
\end{itemize}

\begin{table}[!b]
	\scriptsize
	\centering
	\begin{tabular}{|l|l|l|c|c|c|}
		\hline	
		\textbf{solver} & \textbf{config.} & \textbf{Author} & \textbf{Sub./Sel.} & \textbf{type} & \textbf{ref.}\\
		\hline
		\caqe & - & L.Tentrup, M. Rabe  & selected & CEGAR (clausal abstraction) & \cite{rabe2015caqe}\\	
		& bloqqer &   & & & \\	
		& hqspre &  & & & \\	
		\hline
		\depqbf & - & F. Lonsing & selected & QCDCL & \cite{lonsing2017depqbf}\\
		& no-nenofex &   & & & \\	
		\hline
		\depqbf & cadical & M. Heisinger & selected & QCDCL & \cite{depqbf-cadical}\\
		\hline
		\dynqbf & & G. Charwat & selected & dynamic programming & \cite{charwat2016bdd}\\		
		\hline		
		\hybridqbf & caqe-v1 & L. Zhaohui & submitted & Combination of approaches& \cite{hybridqbf} \\
		& caqe-v2-m1 & & & & \\
		& caqe-v2-m2 & & & & \\
		& caqe-v3 & & & & \\
		& hb-v0 & & & & \\
		& hb-v1 & & & & \\
		& hb-v2 & & & & \\
		& hqsprebloqqer & & & & \\
		\hline
		\miniqu & & F. Slivovsky & selected & QCDCL & \cite{miniqu}\\	
		\hline
		\qute & ofo & F. Slivovsky & submitted & QCDCL & \cite{peitl2019dependency}\\	
		& plain & & & & \\
		& upure & & & & \\
		\hline
		\rareqs & & M. Janota & selected & CEGAR (expansion) & \cite{janota2016solving}\\	
		\hline
	\end{tabular}
	
	\caption{Participating solvers of the PCNF track}
	\label{tbl:pcnf-solver-overview}
\end{table}

\noindent The following commands have been used for calling the solvers:

\begin{lstlisting}
caqe <FILE> 
caqe --preprocessor=bloqqer --qdo <FILE>
caqe --preprocessor=hqspre --qdo <FILE>
depqbf <FILE>
depqbf --qdo --no-dynamic-nenofex <FILE>
depqbf-cadical <FILE>
dynqbf -f <FILE>
HybridQBF_CAQE_V1 <FILE>
HybridQBF_CAQE_V2_M1 <FILE>
HybridQBF_CAQE_V2_M2 <FILE>
HybridQBF_CAQE_V3 <FILE>
HybridQBF_HB_V0 <FILE>
HybridQBF_HB_V1 <FILE>
HybridQBF_HB_V2 <FILE>
HybridQBF_HQSPre_Bloqqer <FILE>
miniQU <FILE>
qute --out-of-order-decisions all <FILE>
qute <FILE>
qute --depscheme upure <FILE>
rareqs <FILE>
\end{lstlisting}

\subsection{PNCNF Solvers}
Table~\ref{tbl:nonpcnf-solver-overview} lists the solvers participating in the PNCNF track.
Only \qute as described in the previous section was newly submitted to this track. 
The remaining solvers were selected from the previous QBF Gallery.

\newcommand{\cqesto}{\texttt{cqesto}\xspace}
\newcommand{\ghostq}{\texttt{GhostQ}\xspace}
\newcommand{\qfun}{\texttt{qfun}\xspace}
\newcommand{\quabs}{\texttt{QuABS}\xspace}

\begin{table}[!b]
	\scriptsize
	\centering
	\begin{tabular}{|l|l|l|c|c|c|}
		\hline	
		\textbf{solver} & \textbf{config.} & \textbf{Author} & \textbf{Sub./Sel.} & \textbf{type} & \textbf{ref.}\\
		\hline
		\cqesto & - & M. Janota & selected & propositional abstraction & \cite{janota2018circuit}\\			
		& cadical & & & & \\
		\hline
		\ghostq & - & W. Klieber & selected & QDPLL& \cite{ghostq}\\			
		& cegar & & & QDPLL + CEGAR & \\
		\hline		
		\miniqu & & F. Slivovsky & selected & QCDCL & \cite{miniqu}\\	
		\hline
		\qfun & - & M. Janota & selected & expansion + ML & \cite{janota2018towards}\\			
		& cadical  & & & & \\
		\hline
		\quabs & & L. Tentrup & selected & CEGAR (literal abstraction) & \cite{tentrup2016solving}\\	
		\hline
		\qute & ofo & F. Slivovsky & submitted & QCDCL & \cite{peitl2019dependency}\\	
		& plain  & & & & \\
		& upure  & & & & \\
		\hline
	\end{tabular}
	
	\caption{Participating solvers of the PNCNF track}
	\label{tbl:nonpcnf-solver-overview}
\end{table}

\noindent In the following we list the different solving paradigms applied by the PNCNF
solvers:

\begin{itemize}
	\item \textbf{Expansion + ML} utilizes machine learning to derive candidate
	strategies during expansion solving~\cite{janota2018towards}.	
	Rather than expanding the formula with individual assignments, learned strategies
	are incorporated. As further samples are collected, strategies 
	as well as the expansion are iteratively refined until a result is obtained.
	
	\item \textbf{Literal abstraction} is a \textbf{CEGAR}-based approach constructing 
	propositional abstractions for quantifier blocks using literals to summarize the 
	interaction between inner and outer blocks~\cite{tentrup2016non}. 
	SAT solving provides candidate assignments and counterexamples, which
	are used to refine the abstractions recursively. 	
	
	\item \textbf{Propositional abstraction} is a circuit-based extension of
	clausal abstraction. 
	Rather than operating on a CNF matrix, the circuit representation 
	of the propositional part of the QBF is preserved during solving.
	As in clausal abstraction, a propositional abstraction is constructed
	for each quantifier level and repeatedly solved and refined using SAT solving~\cite{janota2018circuit}.
	
	\item \textbf{QDPLL} (Davis-Putnam-Logemann-Loveland for 
	QBFs)~\cite{goultiaeva2009beyond} is the foundation of modern QCDCL.
	The search algorithm assigns variables in prefix order while simplifying
	the formula under the current assignment.
	The algorithm can be utilized for non-clausal representation
	and extended with different modalities of learning and propagation --- including 
	\textbf{CEGAR} learning as optionally applied in \ghostq~\cite{ghostq}.	
	
	\item \textbf{QCDCL} is the same paradigm as described for the PCNF solvers.
	PNCNF solvers using QCDCL commonly translate the input internally into a CNF 
	representation by applying a Tseitin-style 
	clausification~\cite{tseitin1983complexity}.
\end{itemize}

\noindent The following commands have been used for calling the solvers:

\begin{lstlisting}
cqesto <FILE> 	
cqesto-cadical <FILE> 
qcir-to-gq.sh <FILE> | ghostq /dev/stdin -s-cnf -q2 -cegar 0
qcir-to-gq.sh <FILE> | ghostq /dev/stdin -s-cnf -q2 -cegar 1
miniQU <FILE>
qfun <FILE>
qfun-cadical <FILE>
quabs <FILE>
qute --out-of-order-decisions all <FILE>
qute <FILE>
qute --depscheme upure <FILE>
\end{lstlisting} 

\newcommand{\dqbdd}{\texttt{DQBDD}\xspace}
\newcommand{\hqs}{\texttt{HQS-2022}\xspace}
\newcommand{\pthree}{\texttt{P3}\xspace}
\newcommand{\pedant}{\texttt{Pedant}\xspace}
\newcommand{\pedantrewrite}{\texttt{Pedant-rewrite}\xspace}

\begin{table}[!b]
	\scriptsize
	\centering
	\begin{tabular}{|l|l|l|c|c|c|}
		\hline	
		\textbf{solver} & \textbf{config.} & \textbf{Author} & \textbf{Sub./Sel.} & \textbf{type} & \textbf{ref.}\\
		\hline
		\dqbdd & & J. S\'{i} & selected & Quantifier Elimination & \cite{Sic2021efficient}\\			
		\hline
		\hqs & & R. Wimmer, C. Scholl & selected & Quantifier Elimination & \cite{gitina2015solving}\\
		\hline		
		\pthree & & F. Reichl, F. Slivovsky & submitted & CEGIS & \cite{p3}\\
		\hline
		\pedant & & F. Reichl, F. Slivovsky & submitted & CEGIS & \cite{reichl2022pedant}\\
		\hline
		\pedantrewrite & - & C. Cheng & submitted & CEGIS & \cite{pedant-rewrite} \\
		& no-res & & & & \\
		\hline
	\end{tabular}	
	\caption{Participating solvers of the DQBF track}
	\label{tbl:dqbf-solver-overview}
\end{table}

\subsection{DQBF}
Table~\ref{tbl:dqbf-solver-overview} lists the solvers participating in the DQBF track.
Three newly submitted solvers were included.
The remaining solvers were selected from the previous QBF Gallery 
2023~\cite{heisinger2026qbfgallery2023}.

\begin{itemize}
	\item \pedant was submitted as the latest version of the established DQBF solver~\cite{reichl2022pedant}.
	\item \pthree is a simplified version of \pedant with optimized internal data-structure for Skolem functions.
	\item \texttt{\pedant-rewrite} is a rewritten version of \pedant submitted in two configurations.
\end{itemize}

\noindent In the following we list the different solving paradigms applied by the DQBF
solvers:

\begin{itemize}	
	\item \textbf{Quantifier elimination} is a DQBF solving approach that
	eliminates variables responsible for the non-linear dependencies
	introduced by the Henkin quantifiers of the formula~\cite{gitina2015solving}.
	This transforms the input into a QBF with linearly
	ordered dependencies, which is then delegated to a QBF solver.
	This technique can also be applied to Binary decision diagram (BDD) like
	datastructures~\cite{Sic2021efficient}.
	
	\item \textbf{CEGIS} (Counter Example Guided Inductive Synthesis) 
	iteratively constructs Skolem functions of a QBF.
	If a candidate function is invalid for some assignment of
	the universal variables, the resulting counterexample is used for 
	refinement by using interpolation based definitions.
	The process continues until correct Skolem functions are obtained or
	the QBF is proven to be false~\cite{reichl2022pedant}.	
\end{itemize}

\noindent The following commands have been used for calling the solvers:
\begin{lstlisting}
dqbdd <FILE>
hqs2 <FILE>
p3 <FILE>
pedant <FILE>
pedant-rewrite <FILE>
pedant-rewrite <FILE> --max-resolutions 1
\end{lstlisting}

\newcommand{\bloqqer}{\texttt{Bloqqer}\xspace}
\newcommand{\preproq}{\texttt{PreProQ}\xspace}

\subsection{Preprocessors}
Table~\ref{tbl:preprocessor-overview} lists the preprocessors participating in the preprocessor track.
Only one new preprocessor, \preproq, was submitted.
In addition, \bloqqer was selected for the PCNF track.

\begin{itemize}
	\item \preproq (Pre-Processor for Q-circuits) is a PNCNF preprocessor that applies circuit-structure 
	simplification techniques to exploit non-CNF properties~\cite{peyrer2024preprocessing}.
\end{itemize}

\begin{table}[!t]
	\scriptsize
	\centering
	\begin{tabular}{|l|l|l|c|c|c|}
		\hline	
		\textbf{processor} & \textbf{format} & \textbf{Author} & \textbf{Sub./Sel.} & \textbf{type} & \textbf{ref.}\\
		\hline
		\bloqqer & PCNF & A. Biere, F. Lonsing, & selected & Clausal simplification & \cite{biere2011bloqqer}\\			
		& & M. Seidl &  & & \\
		\hline		
		\preproq & PNCNF & C. Peyrer, M. Seidl & submitted & Circuit simplification & \cite{peyrer2024preprocessing}\\
		\hline
	\end{tabular}
	\caption{Preprocessors}
	\label{tbl:preprocessor-overview}
\end{table}

\noindent The following parametrization has been used for the preprocessors.
\begin{lstlisting}
bloqqer <FILE>
preproq <FILE>	
\end{lstlisting} 

\section{Evaluation}
This chapter presents the evaluation results of the QBF Gallery 2026.
First we describe the methodology of the \emph{smoke test}.
Then we analyze the various metrics and results from our evaluation
of each track.

\subsection{Smoke Test}
Before the evaluation, we tested all newly available formulas with a timeout of \SI{300}{\second}
using the solvers participating in the complementing track.
The smoke test served three purposes: ensuring that each formula can be parsed without errors,
identifying potential solver issues early and gathering data necessary for the benchmark selection.
A lack of consensus among the results of different solvers may indicates unsound behavior.
In the following, we discuss the relevant incidents.

\newcommand{\qrp}{\texttt{QRP}\xspace}

\begin{itemize}
	\item Failing instance: \texttt{PCNF/connect4/partial7x6/6\_12\_442317624273\_-13\_SAT.qdimacs}\\
	The following table provides an overview of the solver results on this instance:\\
	\begin{minipage}[t]{.4\linewidth}	
		\scriptsize
		\textbf{SAT}\\
		\rule{\textwidth}{0.4pt}
		\caqe\\
		\texttt{\caqe-bloqqer}\\
		\depqbf\\
		\texttt{\depqbf-cadical}\\
		\texttt{\depqbf-no-nenofex}\\
		\texttt{\hybridqbf-hb-v2}\\	
		\texttt{\qute-plain}\\
		\texttt{\qute-ofo}\\
		\rareqs\\
		\miniqu\\
		\texttt{\hybridqbf-caqe-v3}\\
		\texttt{\qute-upure}\\
		\texttt{\hybridqbf-caqe-v2-m1}
	\end{minipage}%
	\begin{minipage}[t]{.4\linewidth}	
		\scriptsize
		\textbf{UNSAT}\\
		\rule{\textwidth}{0.4pt}
		\texttt{\caqe-hqspre}\\
		\texttt{\hybridqbf-caqe-v2-m2}\\
		\texttt{\hybridqbf-caqe-v1}\\
		\texttt{\hybridqbf-hb-v0}\\
		\texttt{\hybridqbf-hb-v1}\\
		\texttt{\hybridqbf-hqsprebloqqer}
	\end{minipage}
	
	\vspace{.3cm}One submitted PCNF formula was reported to be unsatisfiable by six
	solvers, while the other 13 reported satisfiability. 
	We used \depqbf with \qrp certificate generation to verify the result and obtained a successful proof
	of satisfiability.
	Consequently, the six solvers that reported unsatisfiable were disqualified for producing an unsound result.\footnote{\label{fn:hybridqbf}
		After releasing the results, the authors of \hybridqbf informed us that these 
		issues only occur with the \texttt{Release} build of the included preprocessor 
		\texttt{HQSPre} but not using its \texttt{Debug} build.
	}\\
		
	\item Failing instance: \texttt{DQBF/ramsey/general\_R\_3\_3\_04\_sat.dqdimacs}\\
	The following table provides an overview of the solver results on this instance:\\
		\begin{minipage}[t]{.4\linewidth}	
		\scriptsize
		\textbf{SAT}\\
		\rule{\textwidth}{0.4pt}
		\pedant\\
		\pthree\\
		\texttt{\pedantrewrite-no-res}\\
		\pedantrewrite
	\end{minipage}%
	\begin{minipage}[t]{.4\linewidth}	
		\scriptsize
		\textbf{UNSAT}\\
		\rule{\textwidth}{0.4pt}
		\dqbdd\\
		\hqs
	\end{minipage}
	
	\vspace{.3cm}In total, five formulas in the DQBF track showed discrepancies.
	The specified formula is one representative of them.
	In each case, the solvers \dqbdd and \hqs reporting different results.
	To resolve these cases, we used \pedant to extract models from for the respective formulas, 
	which then could be verified using the provided toolchain (\texttt{certifyModel.py}). 
	Consequently, the two solvers which disagreed with the verified results in multiple cases were disqualified.
	
\end{itemize}

\noindent After running the preprocessor track, an additional discrepancy was found for \texttt{\hybridqbf-hb-v2}.\\
The instance \texttt{PCNF/bounded-synthesis/simple\_arbiter\_enc\_6\_\_system\_\_b8.qdimacs}
was reported as unsatisfiable after preprocessing with \bloqqer. 
Consequently, this solver was also disqualified\textsuperscript{\ref{fn:hybridqbf}}.

\begin{figure}[!t]
	\centering
	\includegraphics[width=.8\linewidth]{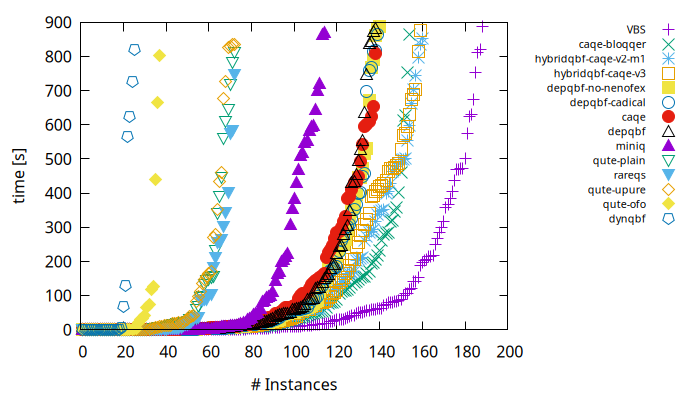}
	\caption{Results of the PCNF track}
	\label{fig:pcnf-result}
\end{figure}

\subsection{PCNF Track}
Table~\ref{tbl:pcnf-evaluation} presents the evaluation results for the PCNF track. 
For completeness, disqualified solvers are listed with an indicator whether their results are counted.
In addition to the participating tools, also the Virtual Best Solver (\texttt{VBS}) is listed.
\texttt{VBS (all)} shows the performance of the Virtual Best solver including the disqualified
solvers.
The solvers are sorted according to their PAR-2 score with a memory limit of \SI{8}{\giga\byte}.
The table additionally reports the number of solved formulas, the distribution of satisfiable and unsatisfiable
formulas, uniquely solved instances and the number of error codes.
Returning \texttt{unknown} (0) was not counted as an error but treated as a timeout
receiving the usual PAR-2 penalty.
For comparison, we also reported PAR-2 scores using a memory limit of \SI{128}{\giga\byte}.
The visualization of those results is shown in Figure~\ref{fig:pcnf-result}.

The winner of the PCNF track is \caqe with \bloqqer as its preprocessor. 
After accounting for disqualifications, \hybridqbf achieved second place, while demonstrating considerable potential
for improved results.
Third place was taken by \depqbf.
Although \dynqbf is ranked last, it uniquely solved two formulas.
Increasing the memory limit did not have a significant effect on the results.

\newcommand{\disqualified}{\rowcolor{red!30}}

\begin{table}[!t]
	\scriptsize
	\centering
	\begin{tabular} {|l|r|r|r|r|r|r|r|}
	\hline
	\textbf{Solver} & \textbf{PAR-2} & \textbf{PAR-2} & \textbf{Solved} & \textbf{sat} & \textbf{unsat} & \textbf{Unique} & \textbf{Errors}\\
	& (\SI{8}{\giga\byte}) & (\SI{128}{\giga\byte}) & & & & & \\
	\hline
	\disqualified VBS (all) & 701.75 & 703.88 & 197 & 82 & 115 & 8 & 0\\
	\hline
	VBS & 746.99 & 747.04 & 188 & 79 & 109 & 10 & 0\\
	\hline
	\disqualified hybridqbf-hb-v1 & 873.23 & 869 & 166 & 71 & 95 & 0 & 0\\
	\hline
	\disqualified hybridqbf-hqsprebloqqer & 885.39 & 889.64 & 163 & 72 & 91 & 1 & 0\\
	\hline
	\disqualified hybridqbf-hb-v2 & 890.29 & 896.07 & 163 & 72 & 91 & 1 & 1\\
	\hline
	\disqualified hybridqbf-hb-v0 & 909.45 & 909.12 & 160 & 70 & 90 & 0 & 0\\
	\hline
	caqe-bloqqer & 929.70 & 929.80 & 154 & 71 & 83 & 0 & 0\\
	\hline
	hybridqbf-caqe-v2-m1 & 949.48 & 949.45 & 152 & 63 & 89 & 0 & 0\\
	\hline
	hybridqbf-caqe-v3 & 960.35 & 960.45 & 151 & 66 & 85 & 3 & 0\\
	\hline
	\disqualified hybridqbf-caqe-v1 & 969.28 & 972.134 & 149 & 67 & 82 & 1 & 0\\
	\hline
	\disqualified caqe-hqspre & 994.75 & 988.924 & 143 & 65 & 78 & 0 & 2\\
	\hline
	\disqualified hybridqbf-caqe-v2-m2 & 1000.51 & 1000.51  & 142 & 64 & 78 & 0 & 0\\
	\hline
	depqbf-no-nenofex & 1019.17 & 1016.41 & 140 & 57 & 83 & 0 & 0\\
	\hline
	depqbf-cadical & 1022.40 & 1022.35 & 139 & 56 & 83 & 0 & 4\\
	\hline
	depqbf & 1030.25 & 1030.40 & 138 & 56 & 82 & 0 & 0\\
	\hline
	caqe & 1030.80 & 1030.89 & 138 & 57 & 81 & 0 & 0\\
	\hline
	miniq & 1168.89 & 1168.76 & 114 & 48 & 66 & 0 & 0\\
	\hline
	rareqs & 1392.47 & 1392.35 & 72 & 37 & 35 & 0 & 0\\
	\hline
	qute-plain & 1399.05 & 1399.02 & 72 & 31 & 41 & 0 & 0\\
	\hline
	qute-upure & 1402.54 & 1402.47 & 72 & 32 & 40 & 0 & 0\\
	\hline
	qute-ofo & 1590.30 & 1590.22 & 37 & 20 & 17 & 0 & 0\\
	\hline
	dynqbf & 1662.56 & 1660.16 & 25 & 16 & 9 & 2 & 0\\
	\hline
\end{tabular}
	\caption{Evaluation data of the PCNF track. Red colored rows are disqualified.}
	\label{tbl:pcnf-evaluation}
\end{table}

\begin{figure}[!b]
	\centering
	\includegraphics[width=.8\linewidth]{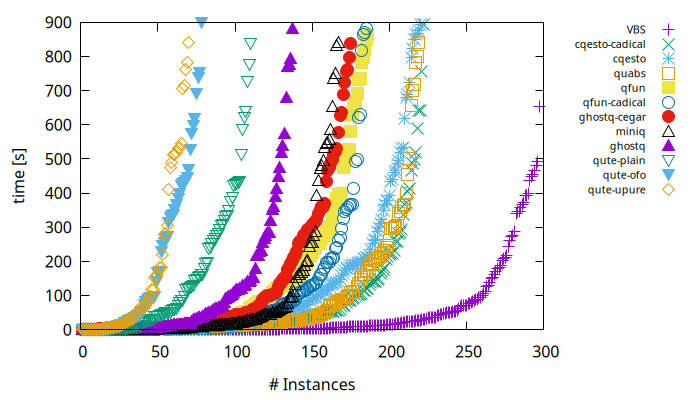}
	\caption{Results of the PNCNF track}
	\label{fig:nonpcnf-result}
\end{figure}

\subsection{PNCNF Track}
Table~\ref{tbl:pncnf-evaluation} presents the evaluation results for the PNCNF track. 
As in the PCNF track, we report the usual statistics as well as the additional evaluation 
with a memory limit of \SI{128}{\giga\byte}.

The winner of this track is \cqesto with an internal \texttt{CaDiCaL} SAT solver. 
Second place was achieved by \quabs, followed by \qfun in third place.
\miniqu is noteworthy for solving the largest number of formulas uniquely.
Increasing the memory limit had a slightly positive effect on some solvers, 
with \cqesto showing the biggest improvement.

\begin{table}[!b]
	\scriptsize
	\centering
\begin{tabular} {|l|r|r|r|r|r|r|r|}
	\hline
	\textbf{Solver} & \textbf{PAR-2} & \textbf{PAR-2} & \textbf{Solved} & \textbf{sat} & \textbf{unsat} & \textbf{Unique} & \textbf{Errors}\\
	& (\SI{8}{\giga\byte}) & (\SI{128}{\giga\byte}) & & & & & \\
	\hline
	VBS & 415.62 & 415.58 & 297 & 168 & 129 & 22 & 0\\
	\hline
	cqesto-cadical & 778.58 & 765.982 & 222 & 133 & 89 & 1 & 0\\
	\hline
	quabs & 796.86 & 793.92 & 219 & 122 & 97 & 6 & 0\\
	\hline
	cqesto & 815.30 & 805.217 & 220 & 123 & 97 & 0 & 0\\
	\hline
	qfun-cadical & 957.02 & 947.965  & 185 & 102 & 83 & 5 & 0\\
	\hline
	qfun & 972.42 & 969.904 & 186 & 101 & 85 & 2 & 0\\
	\hline
	ghostq-cegar & 1019.11 &  1015.93 & 175 & 94 & 81 & 0 & 1\\
	\hline
	miniq & 1037.80 & 1031.22 & 167 & 99 & 68 & 8 & 0\\
	\hline
	ghostq & 1183.00 & 1180.29 & 137 & 83 & 54 & 0 & 1\\
	\hline
	qute-plain & 1314.34 & 1314.35 & 110 & 60 & 50 & 0 & 0\\
	\hline
	qute-ofo & 1463.49 & 1463.47 & 78 & 36 & 42 & 0 & 0\\
	\hline
	qute-upure & 1495.57 & 1495.57 & 70 & 31 & 39 & 0 & 0\\
	\hline
\end{tabular}
	\caption{Evaluation data of the PNCNF track}
	\label{tbl:pncnf-evaluation}
\end{table}

\begin{figure}[!b]
	\centering
	\includegraphics[width=.8\linewidth]{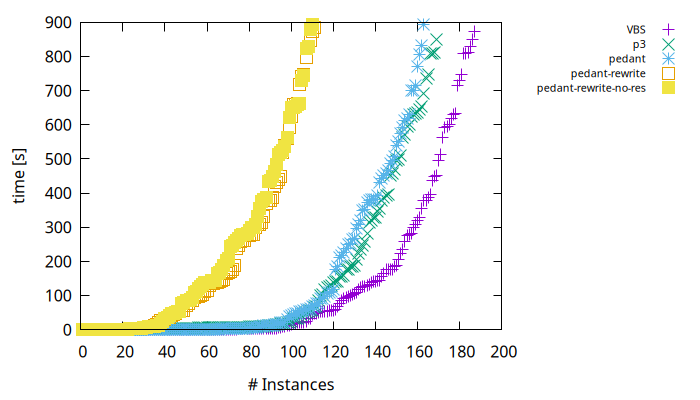}
	\caption{Results of the DQBF track}
	\label{fig:dqbf-result}
\end{figure}

\subsection{DQBF Track}
Table~\ref{tbl:dqbf-evaluation} presents the evaluation for the DQBF track.
As in the previous tracks, the table reports the usual statistics.
Disqualified solvers are listed for completeness and the \texttt{VBS (all)} solver
visualizes the Virtual Best Solver over all participants.

The winner of the DQBF track is \pthree. Second place was achieved by \pedant, followed by \pedantrewrite.  
Even though \dqbdd was disqualified, it is noteworthy for solving the most formulas uniquely. 

\begin{table}[!t]
	\scriptsize
	\centering
	\begin{tabular} {|l|r|r|r|r|r|r|r|}
	\hline
	\textbf{Solver} & \textbf{PAR-2} & \textbf{PAR-2} & \textbf{Solved} & \textbf{sat} & \textbf{unsat} & \textbf{Unique} & \textbf{Errors}\\
	& (\SI{8}{\giga\byte}) & (\SI{128}{\giga\byte}) & & & & &\\
	\hline
	\disqualified VBS (all) & 719.97 & 723.18 & 200 & 91 & 109 & 10 & 0\\
	\hline
	VBS & 821.77 & 821.64 & 187 & 82 & 105 & 23 & 0\\
	\hline	
	p3 & 921.84 & 921.93 & 169 & 67 & 102 & 3 & 0\\
	\hline
	pedant & 956.63 & 950.09 & 161 & 67 & 94 & 0 & 0\\
	\hline
	\disqualified hqs2022 & 1105.16 & 1105.62 & 129 & 61 & 68 & 6 & 1\\
	\hline
	\disqualified dqbdd & 1143.25 & 1143.19 & 120 & 44 & 76 & 1 & 6\\
	\hline
	pedant-rewrite & 1247.71 & 1244.68 & 110 & 58 & 52 & 0 & 0\\
	\hline
	pedant-rewrite-no-res & 1252.27 & 1252.27 & 110 & 59 & 51 & 0 & 0\\
	\hline
\end{tabular}
	\caption{Evaluation data of the DQBF track}
	\label{tbl:dqbf-evaluation}
\end{table}

\subsection{Preprocessor Track}
In this track, we evaluated PCNF and PNCNF tracks by placing a preprocessor before each solver.
The reported runtimes include both preprocessing and solving time.
Table~\ref{tbl:preprocessor-times} presents statistics for preprocessors run without an attached solver.

\begin{table}[!b]
	\centering
	\begin{tabular} {|l|c|c|c|c|c|}
		\hline		
		\textbf{Proc.} & \textbf{Track} & \textbf{total} & \textbf{solved} & \textbf{out} & \textbf{Avg. time}\\
		\hline
		\bloqqer & PCNF & 306 & 34 & 12 & \SI{19.26}{\second}\\
		\hline
		\preproq & PNCNF & 376 & 0 & 1 & \SI{211.62}{\milli\second}\\
		\hline
	\end{tabular}
	\caption{Results of the preprocessors on their corresponding track}
	\label{tbl:preprocessor-times}
\end{table}

\begin{table}[!t]
	\scriptsize
	\centering
	\begin{tabular} {|l|r|r|r|r|r|r|r|}
		\hline
		\textbf{Solver} & \textbf{PAR-2} & \textbf{solved} & \textbf{solved} & \textbf{sat} & \textbf{unsat} & \textbf{Unique} & \textbf{Errors}\\
		(bloqqer+\dots) & & (incl. 34) & (PCNF) & & & & \\
		\hline
		\disqualified VBS (all) & 749.55 & \textbf{202} & 197 & 55 & 113 & 14 & 0\\
		\hline
		VBS & 776.46 & \textbf{197} & 188 & 55 & 108 & 14 & 0\\
		\hline
		depqbf-no-nenofex & 991.07 & \textbf{166} & 140 & 46 & 86 & 1 & 0\\
		\hline
		caqe & 994.66 & \textbf{163} & 138 & 49 & 80 & 0 & 0\\
		\hline
		caqe-bloqqer & 995.558 & \textbf{163} & 154 & 49 & 80 & 0 & 0\\
		\hline	
		hybridqbf-caqe-v2-m1 & 997.09 & \textbf{163} & 152 & 49 & 80 & 0 & 3\\
		\hline
		depqbf-cadical & 1002.79 & \textbf{162} & 139 & 45 & 83 & 0 & 0\\
		\hline
		depqbf & 1022.72 & \textbf{159} & 138 & 43 & 82 & 0 & 0\\
		\hline
		\disqualified hybridqbf-caqe-v1 & 1048.32 & \textbf{156} & 149 & 41 & 81 & 3 & 0\\
		\hline
		\disqualified hybridqbf-hb-v0 & 1049.27 & 156 & \textbf{160} & 42 & 80 & 1 & 40\\
		\hline
		\disqualified hybridqbf-hqsprebloqqer & 1055.68 & 153 & \textbf{163} & 42 & 77 & 0 & 36\\
		\hline
		\disqualified hybridqbf-hb-v1 & 1060.83 & 152 & \textbf{166} & 42 & 76 & 0 & 38\\
		\hline
		\disqualified caqe-hqspre & 1088.49 & \textbf{148} & 143 & 39 & 75 & 0 & 0\\
		\hline
		hybridqbf-caqe-v3 & 1089.29 & 148 & \textbf{151} & 38 & 76 & 1 & 0\\
		\hline
		\disqualified hybridqbf-caqe-v2-m2 & 1092.09 & \textbf{147} & 142 & 38 & 75 & 0 & 3\\
		\hline
		miniq & 1103.37 & \textbf{146} & 114 & 42 & 70 & 2 & 0\\
		\hline
		qute-plain & 1329.66 & \textbf{109} & 72 & 32 & 43 & 1 & 0\\
		\hline
		qute-upure & 1334.80 & \textbf{108} & 72 & 32 & 42 & 1 & 0\\
		\hline
		rareqs & 1387.51 & \textbf{100} & 72 & 29 & 37 & 2 & 0\\
		\hline
		qute-ofo & 1680.42 & \textbf{53} & 37 & 10 & 9 & 0 & 0\\
		\hline
		dynqbf & 1729.95 & \textbf{36} & 25 & 7 & 5 & 2 & 0\\
		\hline					
	\end{tabular}
	\caption{Results of \bloqqer with solvers on the PCNF track}
	\label{tbl:preprocessor-pcnf-times}
\end{table}

Table~\ref{tbl:preprocessor-pcnf-times} presents the results obtained by running \bloqqer
before the solvers in the PCNF track.
We compare the number of formulas solved by each resulting pipeline with the 
corresponding original results from the PCNF track without preprocessing.
Note that the PAR-2 score is computed only over the formulas not trivially solved by the preprocessor,
whereas the number of solved formulas includes these 34 instances.
It is further noticeable that some \hybridqbf configurations suddenly produced
a lot of errors.
These errors originated in the parser, which reported that no input stream was available.
They occurred only sporadically, with no apparent pattern.
We were unable to determine their cause and hope that the authors of the solver can address
this issue in future versions.
The results for \texttt{\hybridqbf-hb-v2} were not processed further because of the 
inconsistency detected directly in this track.
\texttt{\depqbf-no-nenofex} won this track solving 26 additional formulas.
Second place went to \caqe, which solved 25 additional instances followed by
\hybridqbf in third place despite three erroneous instances.
\qute benefited the most from \bloqqer, solving 37 additional formulas.
 
Table~\ref{tbl:preprocessor-pncnf-times} presents the results of running \preproq before the
solvers in the PNCNF track.
Due to the clausification capabilities of this preprocessor, we additionally included the PCNF
track winner \caqe. 
The instance \texttt{PNCNF/Gallery23/hyperprop/2.2.qcir} caused \preproq to exceed
its \SI{8}{\giga\byte} memory limit.
This subsequently resulted in an error for the attached solver, explaining the single
error reported for every solver in this track

The winner of this track is \texttt{\cqesto-cadical}. 
Notably, the base version of \cqesto did not benefit to the same extent.
Second place was achieved by \quabs, followed by \miniqu in third place.
The biggest benefit was observed for \texttt{\qute-ofo}, which solved 20 additional instances.

\begin{table}[!t]
	\scriptsize
	\centering
	\begin{tabular} {|l|r|r|r|r|r|r|r|}
	\hline
	\textbf{Solver} & \textbf{PAR-2} & \textbf{solved} & \textbf{solved} & \textbf{sat} & \textbf{unsat} & \textbf{Unique} & \textbf{Errors}\\
	(preproq+\dots) & & & (PNCNF) & & & & \\
	\hline
	VBS & 415.22 & 297 & 297 & 167 & 130 & 20 & 1\\
	\hline
	cqesto-cadical & 701.43 & \textbf{239} & 222 & 140 & 99 & 3 & 1\\
	\hline
	quabs & 791.87 & \textbf{220} & 219 & 127 & 93 & 7 & 1\\
	\hline
	cqesto & 841.08 & 213 & \textbf{220} & 118 & 95 & 0 & 1\\
	\hline
	miniq & 962.52 & \textbf{181} & 167 & 105 & 76 & 7 & 1\\
	\hline
	qfun-cadical & 969.99 & 184 & \textbf{185} & 103 & 81 & 2 & 1\\
	\hline
	qfun & 1015.23 & 177 & \textbf{186} & 98 & 79 & 1 & 1\\
	\hline
	ghostq-cegar & 1050.73 & 167 & \textbf{175} & 92 & 75 & 0 & 1\\
	\hline
	ghostq & 1184.33  & \textbf{138} & 137 & 84 & 54 & 0 & 1\\
	\hline
	qute-plain & 1253.48 & \textbf{122} & 110 & 63 & 59 & 0 & 1\\
	\hline
	qute-ofo & 1354.45 & \textbf{98} & 78 & 49 & 49 & 0 & 1\\
	\hline
	caqe & 1383.04 &  \textbf{93} & - & 40 & 53 & 0 & 1\\
	\hline
	qute-upure & 1407.87 & \textbf{89} & 70 & 36 & 53 & 0 & 1\\
	\hline
\end{tabular}
\caption{Results of \preproq with solvers on the PNCNF track}
\label{tbl:preprocessor-pncnf-times}
\end{table}

\begin{table}[!t]
	\centering
	\resizebox{\textwidth}{!}{
		\begin{tabular} {|l|c|c|c|c|c|c|c|c|c|c|c|c|c|c|c|c|}
			\hline
			\textbf{Solver} & \textbf{F0} & \textbf{F1} & \textbf{F2} & \textbf{F3} & \textbf{F4} & \textbf{F5} & \textbf{F6} & \textbf{F7} & \textbf{F8} & \textbf{F9} & \textbf{F10} & \textbf{F11} & \textbf{F12} & \textbf{F13} & \textbf{F14} & \textbf{F15}\\
			\hline
			caqe & 12 & 1 & 14 & 12 & 14 & 13 & 15 & 13 & 10 & \checkmark & \checkmark & \checkmark & \checkmark & \checkmark & \checkmark & \checkmark\\
			\hline
			caqe-bloqqer & 13 & 3 & \checkmark & \checkmark & \checkmark & \checkmark & \checkmark & \checkmark & 16 & \checkmark & \checkmark & \checkmark & \checkmark & \checkmark & \checkmark & \checkmark\\
			\hline
			caqe-hqspre & 12 & \checkmark & 14 & 13 & 15 & 13 & 15 & 13 & 14 & \checkmark & \checkmark & \checkmark & \checkmark & 14 & \checkmark & 30\\
			\hline
			depqbf & 19 & 1 & 50 & 12 & \checkmark & \checkmark & \checkmark & \checkmark & 10 & \checkmark & \checkmark & \checkmark & \checkmark & \checkmark & \checkmark & \checkmark\\
			\hline
			depqbf-cadical & 18 & 1 & 50 & 12 & \checkmark & \checkmark & \checkmark & \checkmark & 10 & \checkmark & \checkmark & \checkmark & \checkmark & \checkmark & \checkmark & \checkmark\\
			\hline
			depqbf-no-nenofex & 21 & 1 & 20 & 12 & 24 & 12 & 24 & 12 & 10 & \checkmark & \checkmark & \checkmark & \checkmark & \checkmark & \checkmark & \checkmark\\
			\hline
			dynqbf & \checkmark & 1 & \checkmark & \checkmark & \checkmark & \checkmark & \checkmark & \checkmark & 30 & \checkmark & \checkmark & \checkmark & \checkmark & \checkmark & \checkmark & \checkmark\\
			\hline
			hybridqbf-caqe-v1 & 13 & \checkmark & 14 & 13 & 15 & 13 & 15 & 13 & 14 & \checkmark & \checkmark & \checkmark & \checkmark & 24 & \checkmark & 0\\
			\hline
			hybridqbf-caqe-v2-m1 & 13 & 3 & 14 & 12 & 14 & 13 & 15 & 13 & 10 & \checkmark & \checkmark & \checkmark & \checkmark & \checkmark & \checkmark & \checkmark\\
			\hline
			hybridqbf-caqe-v2-m2 & 12 & \checkmark & 14 & 13 & 15 & 13 & 15 & 13 & 14 & \checkmark & \checkmark & \checkmark & \checkmark & 40 & \checkmark & 0\\
			\hline
			hybridqbf-caqe-v3 & 14 & 3 & 14 & 13 & 15 & 13 & 15 & 13 & 15 & \checkmark & \checkmark & \checkmark & \checkmark & \checkmark & \checkmark & \checkmark\\
			\hline
			hybridqbf-hb-v0 & 14 & 2 & 24 & 13 & 24 & 13 & 23 & 13 & 19 & \checkmark & \checkmark & \checkmark & \checkmark & \checkmark & \checkmark & \checkmark\\
			\hline
			hybridqbf-hb-v1 & 14 & 2 & 24 & 13 & 24 & 13 & 23 & 13 & 19 & \checkmark & \checkmark & \checkmark & \checkmark & \checkmark & \checkmark & \checkmark\\
			\hline
			hybridqbf-hb-v2 & 14 & 2 & 24 & 13 & 24 & 13 & 23 & 13 & 20 & \checkmark & \checkmark & \checkmark & \checkmark & \checkmark & \checkmark & \checkmark\\
			\hline
			hybridqbf-hqsprebloqqer & 14 & 2 & 24 & 13 & 24 & 13 & 23 & 13 & 19 & \checkmark & \checkmark & \checkmark & \checkmark & \checkmark & \checkmark & \checkmark\\
			\hline
			miniq & 22 & 1 & 24 & 12 & 23 & 12 & 24 & 12 & 11 & \checkmark & \checkmark & \checkmark & \checkmark & \checkmark & \checkmark & \checkmark\\
			\hline
			qute-ofo & \checkmark & 1 & \checkmark & \checkmark & \checkmark & \checkmark & \checkmark & \checkmark & 12 & \checkmark & \checkmark & 5 & 9 & 0 & 0 & 0\\
			\hline
			qute-plain & \checkmark & 1 & \checkmark & \checkmark & \checkmark & \checkmark & \checkmark & \checkmark & 12 & \checkmark & \checkmark & \checkmark & \checkmark & \checkmark & \checkmark & 23\\
			\hline
			qute-upure & \checkmark & 1 & \checkmark & \checkmark & \checkmark & \checkmark & \checkmark & \checkmark & 12 & \checkmark & \checkmark & \checkmark & \checkmark & \checkmark & \checkmark & 16\\
			\hline
			rareqs & 11 & 1 & 14 & 12 & 14 & 11 & 14 & 11 & 10 & \checkmark & \checkmark & \checkmark & \checkmark & \checkmark & \checkmark & \checkmark\\
			\hline
		\end{tabular}
	}
	\caption{Crafted formulas}
	\label{tbl:crafted-results}	
\end{table}

\subsection{Crafted Instances Track}
Table~\ref{tbl:crafted-results} presents the results of the Crafted Instances track.
Due to space limitations, we assigned aliases to the instances.
The following list gives the corresponding mappings:

\begin{minipage}{.5\linewidth}
\scriptsize
F0=\texttt{AESAT}(n)\\
F1=\texttt{CLIQUECOL}(m)\\
F2=\texttt{GADGET}(n, base=ec, gadget=eq)\\
F3=\texttt{GADGET}(n, base=ec, gadget=xor)\\
F4=\texttt{GADGET}(n, base=ic, gadget=eq)\\
F5=\texttt{GADGET}(n, base=ic, gadget=xor)\\
F6=\texttt{GADGET}(n, base=sc, gadget=eq)\\
F7=\texttt{GADGET}(n, base=sc, gadget=xor)\\
\end{minipage}%
\begin{minipage}{.5\linewidth}
\scriptsize
F8=\texttt{SUBSETSUM}(n)\\
F9=\texttt{SUCCKRAD}(n, k=3)\\
F10=\texttt{SUCCKRAD}(n, k=5)\\
F11=\texttt{SUCCKRAD}(n, k=7)\\
F12=\texttt{SUCCKRAD}(n, k=10)\\
F13=\texttt{SUCCKRAD}(n, k=15)\\
F14=\texttt{SUCCKRAD}(n, k=20)\\
F15=\texttt{SUCCKRAD}(n, k=30)
\end{minipage}
\noindent with $n = \{5 \dots 30\} \cup \{40, 50, 60 ,70\}$ and
$m = \{1 \dots 4\}$.

\noindent In this track, the parameter $n$ is increased incrementally until either
the time limit is reached or the final instance is solved. 
Each table entry denotes the largest $n$ solved successfully within the resource limitations. 
A \checkmark\xspace indicates that all instances of the respective family were solved successfully.

The \texttt{CLIQUECOL} instances were the most challenging in this track
as they were the only family for which the range of $n$ had to be adjusted.
Even on the range of $n=1\dots4$, the fastest runtime was still above \SI{500}{\second}.
In contrast, the \texttt{SUCKRAD} instances were the easiest to solve.
No solver solved all \texttt{SUBSETSUM} instances up to $n=70$.
Overall, \dynqbf performed the best in this track, successfully solving every family
except \texttt{CLIQUECOL} and \texttt{SUBSETSUM}.

\section{Beyond PAR-2 Scores}
This section presents additional statistics for the three main tracks.
In particular, we examine similarities between solvers and analyze the performance of each solver
on the individual formula families.
To this end we visualize solver similarities using heatmaps, following a concept
introduced at the SAT Competition 2020~\cite{froleyks2021sat}.
For each pair of solvers, we compute the Pearson correlation coefficient~\cite{pearson1895note} of
their runtimes over the individual instances.
The resulting values are represented by a color gradient ranging from green (strong inverse correlation) 
to yellow (no correlation) to red (strong correlation).
A solver compared with itself therefore appears in red on the diagonal of the heatmap.
Furthermore, we provide histograms showing the results for each participating solver, 
segmented by the formula families.

\begin{figure}[!t]
	\centering
	\includegraphics[width=\linewidth]{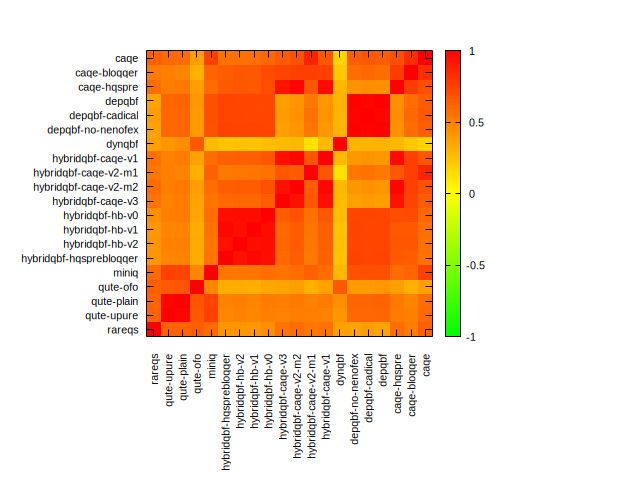}
	\caption{Solver similarities of the PCNF track visualized in a heatmap}
	\label{fig:pcnf-heatmap}
\end{figure}

\begin{figure}[!b]
	\centering
	\includegraphics[width=\linewidth]{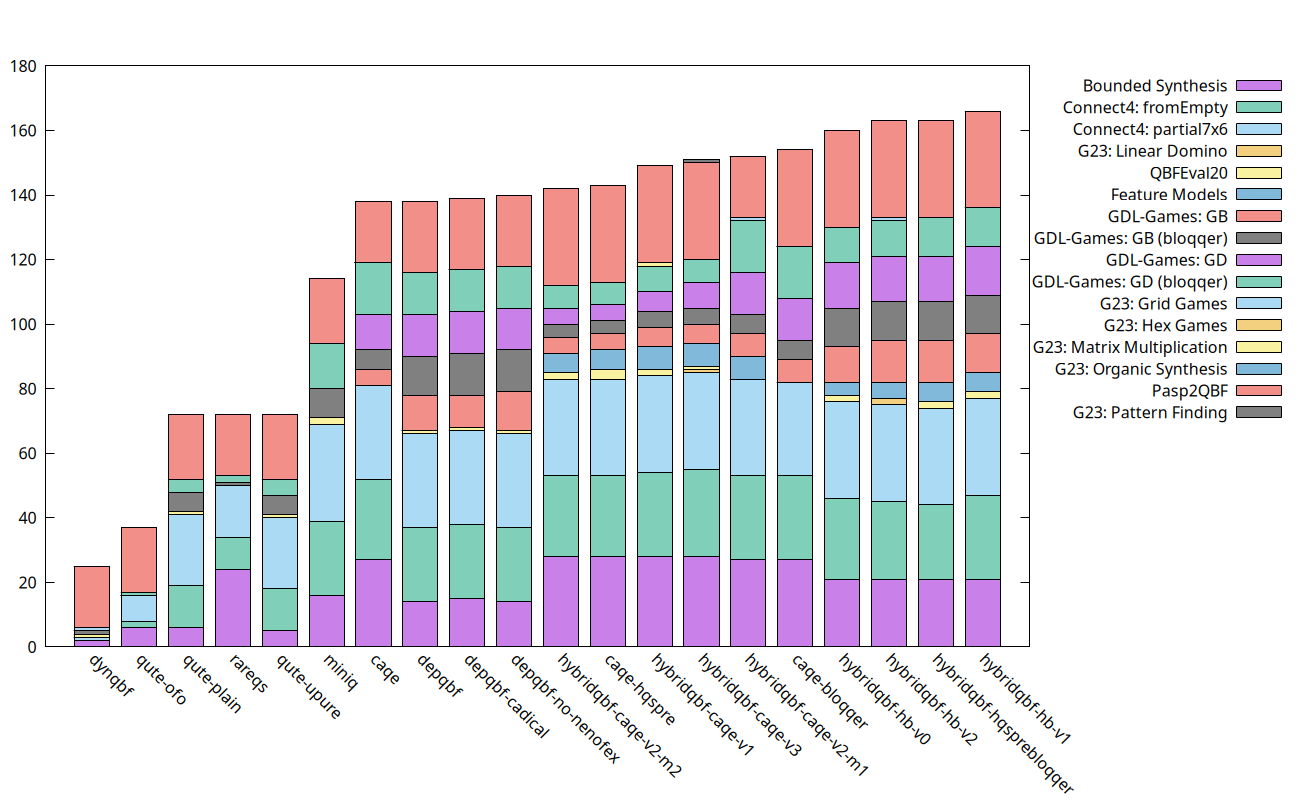}
	\caption{Formula families solved by the solvers of the PCNF track}
	\label{fig:pcnf-hist}
\end{figure}

\begin{figure}[!t]
	\centering
	\includegraphics[width=.8\linewidth]{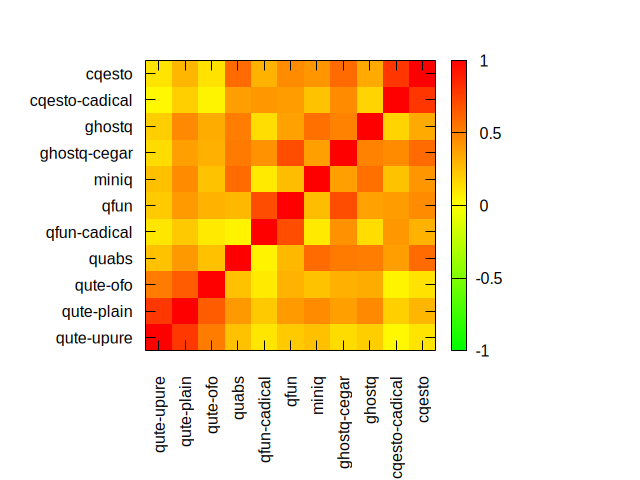}
	\caption{Solver similarities of the PNCNF track visualized in a heatmap}
	\label{fig:nonpcnf-heatmap}
\end{figure}

\begin{figure}[!t]
	\centering
	\includegraphics[width=\linewidth]{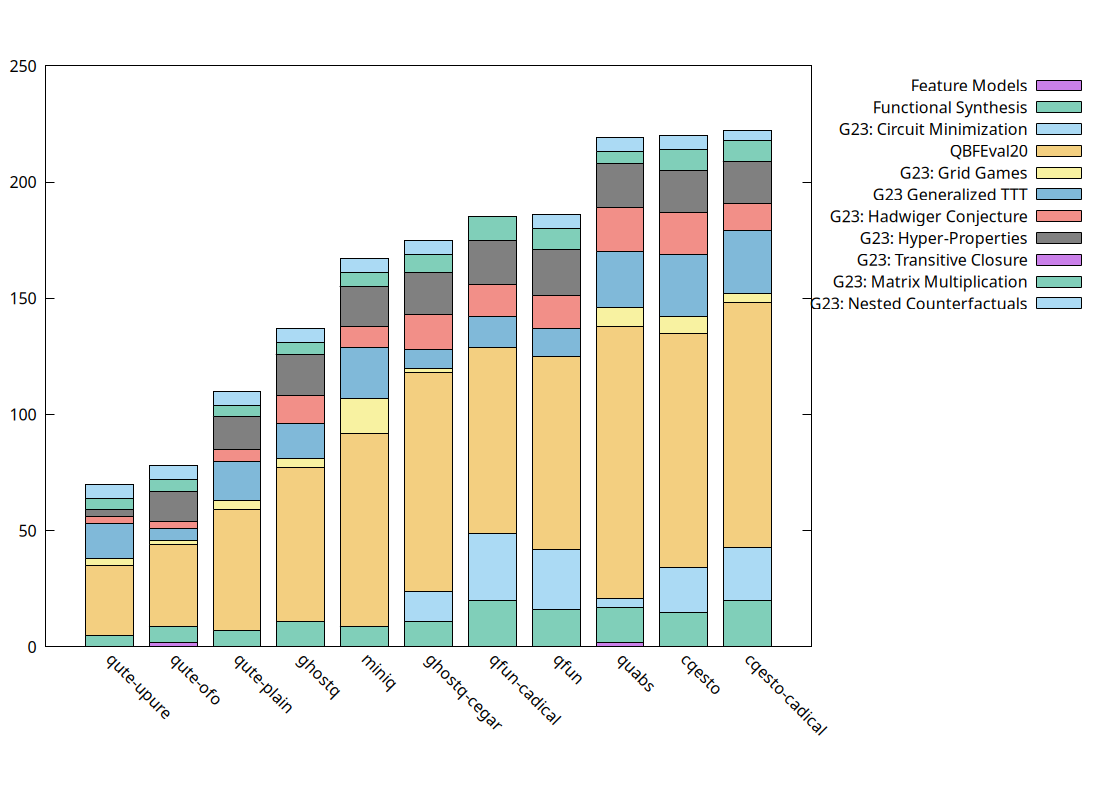}
	\caption{Formula families solved by the solvers of the PNCNF track}
	\label{fig:nonpcnf-hist}
\end{figure}

\subsection{PCNF Track}
Figure~\ref{fig:pcnf-heatmap} shows the similarities between the solvers in the PCNF track.
The solvers least similar to the other participants were \dynqbf and \qute with out-of-order decisions.
\hybridqbf and its \texttt{hb} variants were the most similar to one another. 
Overall, the track exhibited a moderate degree of variation between solver behaviors.
Figure~\ref{fig:pcnf-hist} shows the number of solved formulas for each solver, segmented by formula family.
The \texttt{Pasp2QBF} family was solved the most frequently overall, whereas the solvers 
struggled the most with the \texttt{Gallery 23:} \texttt{Pattern finding} instances.

The \texttt{Feature model} family has by far the largest average number of
clauses and variables, whereas \texttt{Gallery23:\text{ }Grid Games} has the smallest
averages for both measures. Both \texttt{Pasp2QBF} and \texttt{Feature model}
contain only one quantifier alternation, while \texttt{GDL-Games:} \texttt{GB} has the
largest average number of quantifier alternations. 
The hard to solve \texttt{Gallery23:} \texttt{Pattern finding} family 
does not stand out in any of these metrics.
Overall, no strong correlation between those statistics and the hardness
of a formula was observed.

\subsection{PNCNF Track}
Figure~\ref{fig:nonpcnf-heatmap} shows the similarities between the solvers in the PNCNF track.
The variation among these solvers is substantially higher than in the PCNF track, indicating less similar 
solver behavior overall.
The most similar solvers are primarily different configurations of the same solver. 
Figure~\ref{fig:nonpcnf-hist} shows the number of solved formulas for each solver, segmented by formula family.
The highest number of solved instances comes from the QBFEval2020, which unsurprising since this family accounts for
approximately \SI{50}{\percent} of the instances of the track. 
The most challenging families were the new \texttt{Feature Models} and the \texttt{Gallery23:} 
\texttt{Transitive Closure}.

The \texttt{Gallery23:} \texttt{Hyper-Properties} family has the largest average number
of gates, while having a comparatively moderate average number of variables and 
the \texttt{Feature Models} family has the largest average number of variables.
\texttt{Generalized TTT} has the smallest average numbers of both gates
and variables. 
The difficult \texttt{Gallery23:} \texttt{Transitive Closure} family has
the second-largest average number of variables. 
Overall, the results indicate a slight trend towards increased 
solving time for families with more variables.

\begin{figure}[!b]
	\centering
	\includegraphics[width=.7\linewidth]{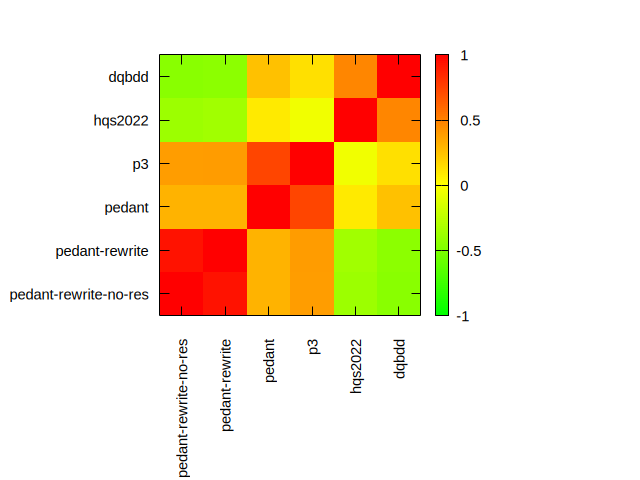}
	\caption{Solver similarities of the DQBF track visualized in a heatmap}
	\label{fig:dqbf-heatmap}
\end{figure}

\begin{figure}[!t]
	\centering
	\includegraphics[width=.9\linewidth]{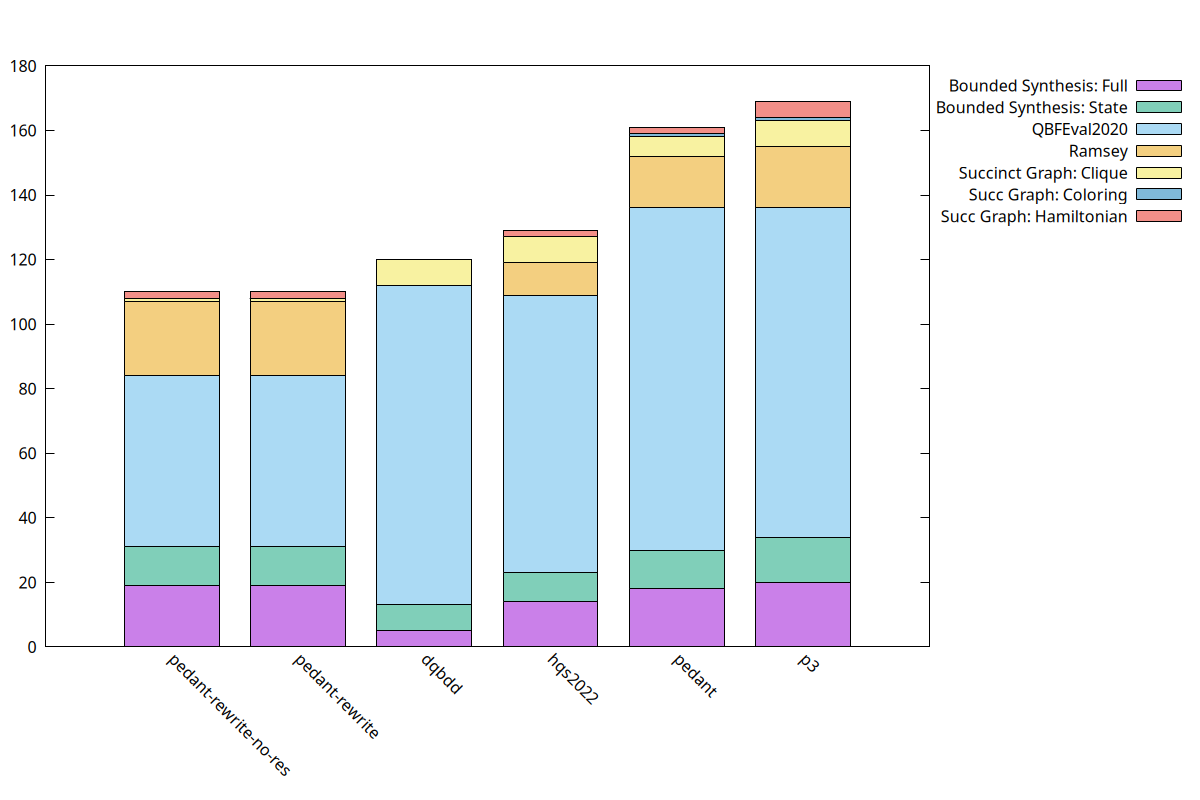}
	\caption{Formula families solved by the solvers of the DQBF track}
	\label{fig:dqbf-hist}
\end{figure}

\subsection{DQBF Track}
Figure~\ref{fig:dqbf-heatmap} shows the similarities between the solvers of the DQBF track. 
\texttt{\pedant-rewrite}, \dqbdd and \hqs differ most from the remaining solvers in this track. 
\pedant and \pthree are the most similar.

Figure~\ref{fig:dqbf-hist} shows the number of solved formulas for each solver, segmented by formula family.
Most solved instances originated from the \texttt{QBFEval2020} family, which accounts for more than 
half of the formulas in the track.
The \texttt{Succinct Graph:} \texttt{Coloring} family was the most challenging overall.

The \texttt{Bounded Synthesis:} \texttt{Full} family has by far the largest average
numbers of clauses and variables, while also having the lowest proportion of
variables quantified by Henkin quantifiers at only \SI{0.014}{\percent}. 
The \texttt{Succinct Graph:} \texttt{Coloring} family has the smallest average numbers of
both variables and clauses. 
In contrast, \texttt{Succinct Graph:} \texttt{Clique} has the highest average proportion of 
Henkin-quantified variables at \SI{27.53}{\percent}. 
Overall, these metrics do not appear to correlate with formula hardness.

\section{Conclusion}
The QBF Gallery 2026 continued the traditional series of QBF evaluations. 
We evaluated a range of new formulas and solvers with a noticeable increase in interest in DQBF.
All submitted instances and solvers were described, analyzed and tested
before conducting the detailed evaluation.

In addition to the main tracks, we evaluated new crafted instances and investigated the effect of preprocessing.
Beyond the usual PAR-2 scores, we also provided insights into solver similarities and performance on
individual formula families.
Finally, we released the new benchmarks in a dedicated artifact~\cite{gallery26-instances}.

%%
%% Bibliography
%%

%% Please use bibtex, 

\bibliography{sources}

\appendix
\newpage
\section{Results with disqualified solvers}
\begin{figure}[!h]
	\centering
	\includegraphics[width=.9\linewidth]{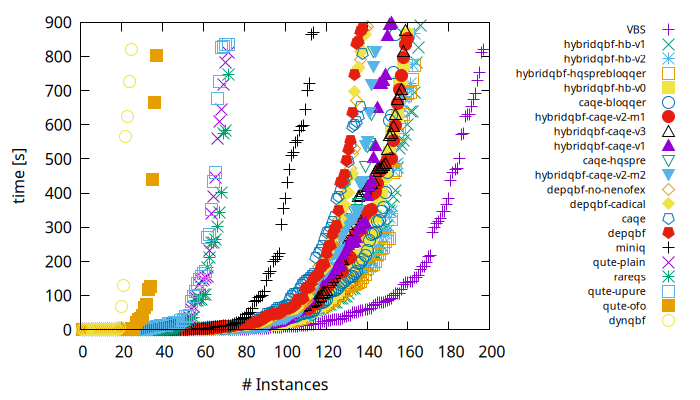}
	\caption{Results of the PCNF track containing also the disqualified solvers}
	\label{fig:pcnf-result-all}
\end{figure}

\begin{figure}[!h]
	\centering
	\includegraphics[width=.9\linewidth]{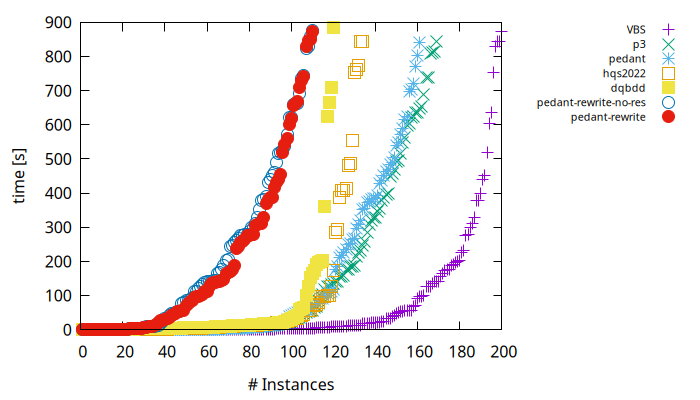}
	\caption{Results of the DQBF track containing also the disqualified solvers}
	\label{fig:dqbf-result-all}
\end{figure}

Figure~\ref{fig:pcnf-result-all} and \ref{fig:dqbf-result-all} show the results of the PCNF and DQBF
track with the disqualified solvers. 

\end{document}